\documentclass[a4paper,11pt]{article}

\usepackage[utf8]{inputenc}
\usepackage{amsmath,amssymb}
\usepackage{graphicx}
\usepackage{geometry}
\usepackage{array}
\usepackage[table,xcdraw]{xcolor}
\usepackage{multicol}
\usepackage{tabularx}
\usepackage{subcaption}
\usepackage{jhep}
\usepackage{physics}
\usepackage{float}
\usepackage{tikz}
\usepackage{tikz-feynman}
\tikzfeynmanset{compat=1.1.0}
\usepackage{mathrsfs}

\title{\boldmath Modular $S_4$ Scotogenic Model with Flavored Resonant Leptogenesis}

\author{Abhishek\textsuperscript{1,*}, V. Suryanarayana Mummidi\textsuperscript{1,$\dagger$}}
\affiliation{\textsuperscript{1}Department of Physics, National Institute of Technology,\\
Tiruchirappalli-620015, India}

\emailAdd{413120051@nitt.edu,venkata@nitt.edu}

\abstract{
\begin{abstract}
We construct a radiative neutrino mass model that combines the scotogenic 
mechanism with modular $S_4$ flavour symmetry. The entire lepton flavour 
structure is governed by holomorphic modular forms of a single complex 
modulus $\tau$, eliminating the need for flavon fields. Beyond the Standard 
Model, the particle content consists of two right-handed Majorana fermions 
assigned to the $S_4$ doublet representation and an inert scalar doublet 
odd under a $\mathbb{Z}_2$ parity. Neutrino masses emerge at one loop 
through the scotogenic mechanism, and the lightest $\mathbb{Z}_2$-odd 
state serves as a dark matter candidate. A comprehensive scan of the 
parameter space demonstrates consistency with all five neutrino oscillation 
observables at the $3\sigma$ level. Having exactly two right-handed 
neutrinos forces the light neutrino mass matrix to rank two, leaving one 
neutrino massless and selecting normal ordering as the only viable option. 
The framework predicts a total neutrino mass in the narrow window 
$\Sigma m_\nu \simeq 0.059$--$0.06\,\mathrm{eV}$, well within current 
cosmological bounds, and an effective Majorana mass 
$m_{\beta\beta} \simeq (1.3$--$3.5)\times 10^{-3}\,\mathrm{eV}$ 
relevant for neutrinoless double beta decay searches. The modular structure 
of the right-handed Majorana mass matrix intrinsically produces a 
quasi-degenerate heavy neutrino spectrum, enabling flavoured resonant 
leptogenesis at $M_1 \sim 10^5\,\mathrm{GeV}$ without any fine-tuning. 
Integration of the full three-flavour Boltzmann equations confirms that the 
observed baryon asymmetry is reproduced, establishing that neutrino masses, 
leptonic mixing, and the baryon asymmetry of the Universe all 
find a common explanation within this framework.
\end{abstract}
}

\begin{document}
\maketitle
\newpage
\section{Introduction}

Neutrinos have a tiny mass and the three active flavor states mix at large angles, according to a convincing body of experimental evidence gathered over the previous three decades.
Precision measurements from solar, atmospheric, and reactor experiments have pinned down both independent mass-squared splittings and all three leptonic mixing angles~\cite{SNO:2001kpb,SuperK:1998jle,KamLAND:2008eeq,Maki:1962mu}. These measurements indicate an underlying flavor structure described by the Pontecorvo–Maki–Nakagawa–Sakata (PMNS) matrix. Despite this progress, a number of 
fundamental questions remain unanswered: such as how neutrino masses are ordered, what sets their overall scale, and how large CP violation is in the lepton sector and crucially, the 
underlying dynamical mechanism through which neutrinos acquire mass. 
Since the renormalisable Standard Model (SM) with purely left-handed 
neutrino fields predicts strictly massless neutrinos, addressing these 
questions demands new physics beyond the SM.

Radiative mass generation offers a theoretically attractive path 
forward. Rather than introducing a mass term at tree level, neutrino 
masses arise as calculable loop corrections, with their smallness 
relative to the electroweak scale explained naturally by the loop 
suppression factor. The scotogenic model of Ma~\cite{Ma:2006km} offers a minimal realisation of this idea: two or more right-handed singlet fermions and an inert scalar doublet are added to the SM, both odd under a discrete $\mathbb{Z}_2$ symmetry. This parity forbids a tree-level Dirac mass term and stabilises the lightest $\mathbb{Z}_2$-odd state, providing a natural dark matter (DM) candidate. The framework therefore addresses two fundamental open problems—the origin of neutrino masses and the nature of dark matter—within a single minimal framework.

A limitation of the scotogenic model in its original form is that it imposes no structure on the Yukawa 
matrices and therefore makes no prediction for the observed pattern of 
mixing angles. Explaining the large atmospheric and solar mixing angles 
alongside the comparatively small reactor angle requires additional 
theoretical input. Groups such as $A_4$, $S_4$, and $A_5$ have been extensively used to constrain Yukawa textures~\cite{Ishimori:2010au, Nguyen:2021yuy, Ganguly:2022fkx, Behera:2020yfj, Vien:2023ovr, Ma:2005ke, Ma:2015xla, Ma:2004zv, Bazzocchi:2012pp, Vien:2020qmt, Thapa:2021lga, Ma:2005re, Dev:2015bna, Grossman:2014kya, Petcov:2026abc, Abhishek:2025ijv}, but conventional implementations require flavon fields whose vacuum expectation values (VEVs) break the symmetry, introducing numerous additional parameters and a dependence on the details of the flavon potential.

Modular flavor symmetry provides a better approach to this problem~\cite{Feruglio(2018),Kobayashi:2019, Penedo:2019, Wang:2020, Abhishek:2025ety, Abhishek:2026hex}. Modular invariant models promote Yukawa couplings to holomorphic modular forms of $\tau$, each carrying a definite representation of the finite modular group, so symmetry alone fixes the coupling structure. The entire lepton flavor pattern is therefore encoded in the single complex parameter $\tau$, without the requirement of a flavon sector. As a result, the number of independent parameters decreases substantially compared to traditional discrete flavor models, and the predictive power of the framework is significantly increased. The masses and mixings of quark and lepton have been successfully described using phenomenologically relevant finite modular groups $\Gamma_2 \simeq S_3$, $\Gamma_3 \simeq A_4$, $\Gamma_4 \simeq S_4$, and $\Gamma_5 \simeq A_5$.

A further cosmological puzzle that any complete theory of neutrino 
mass must eventually address is the observed baryon
asymmetry of the Universe (BAU). The baryon-to-photon ratio of $Y_B^{\text{obs}} \simeq 8.6 \times 10^{-11}$~ \cite{Planck:2018vyg,Davidson:2008bu} is determined by precise measurements of the cosmic microwave background radiation and light-element abundances from primordial nucleosynthesis. This value cannot be produced from symmetric initial conditions without meeting the three Sakharov criteria~\cite{Sakharov:1967dj}. One of the more convincing methods for producing this asymmetry is leptogenesis ~\cite{Davidson:2008bu}. 
In the early Universe, CP-violating out-of-equilibrium decays of heavy Majorana neutrinos result in a net lepton number, which is then partially transformed into a baryon number by non-perturbative sphaleron transitions \cite{Buchmuller:2005eh, Fukugita:1986hr}.

The right-handed neutrinos (RHNs) that produce radiative mass also act as an origin of the lepton asymmetry in the scotogenic model~\cite{Tapender:2026ets}, directly connecting two phenomenologically different parts of the theory. The standard hierarchical leptogenesis scenario, however, is known to be inadequate at low mass scales with only two RHNs: a Davidson--Ibarra-type bound on the CP asymmetry necessitates $M_1 \gtrsim 10^{9}$--$10^{10}$~GeV for successful leptogenesis in the scotogenic framework~\cite{Hugle_2018}, making low-scale baryogenesis impossible without extra structure. However, the self-energy contribution to the decay asymmetry experiences a resonant enhancement when the heavy neutrino spectrum displays a near-degeneracy, allowing successful baryogenesis at neutrino mass scales as low as the TeV range—a situation known as resonant leptogenesis~\cite{Pilaftsis:2003gt, Dev:2017wwc}. This near-degeneracy automatically results from the modular nature of the right-handed Majorana mass matrix, without any fine-tuning of the heavy neutrino masses, which is a very appealing aspect of the current model. Furthermore, the Yukawa interactions of the charged leptons maintain thermal equilibrium when leptogenesis progresses at temperatures below about $10^9$~GeV, resulting in the evolution of the different lepton flavors as distinct dynamical entities. In this regime, a fully 
flavoured treatment of the Boltzmann equations is mandatory for obtaining reliable predictions for $Y_B$~\cite{Samanta:2019yeg}.

Motivated by the above considerations, we propose a modular-invariant scotogenic model with modular $S_4$ symmetry. Within this framework, the particle content and symmetry assignments lead to a rank-two light neutrino mass matrix, predicting one massless neutrino state and favoring the normal mass ordering. All Yukawa couplings in the model are expressed as modular 
forms of $S_4$, eliminating the need for any flavon fields and reducing 
the parameter count significantly compared with conventional 
approaches. The model's $S_4$ modular structure of the right-handed Majorana mass matrix naturally induces a quasi-degenerate heavy neutrino spectrum without the need for artificial fine-tuning, which is a crucial component for resonant CP asymmetry enhancement during leptogenesis.

In the $3\sigma$ intervals of the NuFIT 5.2 global fit~\cite{Esteban:2024vvu}, we conduct a comprehensive numerical scan over the complex modulus $\tau$ and the remaining Yukawa parameters, choosing only those points that reproduce each of the five observed neutrino oscillation observables. Within the 
resulting viable parameter space, we assess the prospects for 
flavoured resonant leptogenesis by integrating the complete 
three-flavour Boltzmann equation system for representative benchmark 
configurations and comparing the predicted baryon yield with the 
cosmological value.

This paper's remaining sections are arranged as follows.
The model's particle composition, symmetry assignments, Yukawa Lagrangian, scalar potential, and relevant mass matrices are all introduced in Section~\ref{sec:2}.
The parameter scan results and neutrino phenomenology are presented in Section~\ref{sec:3}.
The analysis of flavored resonant leptogenesis is covered in Section~\ref{sec:4}, along with the Boltzmann-equation evolution for a few benchmark points. In Section~\ref{sec:6}, we conclude our main findings and discuss future research directions.

\section{The Model}
\label{sec:2}
\subsection{Modular Symmetry and the Fundamental Domain}

The starting point is the modular group $\Gamma$ acting on the upper half-plane
$\mathbb{H}=\{\tau\in\mathbb{C}\,|\,\mathrm{Im}\,\tau>0\}$ via M\"obius transformations~\cite{Feruglio(2018),Novichkov:2018yse,Penedo:2019}
\begin{equation}
    \tau \;\longrightarrow\; \gamma\tau \;=\; \frac{a\tau + b}{c\tau + d}\,,
    \qquad a,b,c,d \in \mathbb{Z}\,,\quad ad - bc = 1\,,
    \label{eq:modular_trans}
\end{equation}
Since $(a,b,c,d)$ and its negative yield the same transformation, the group is effectively $\Gamma \cong \mathrm{PSL}(2,\mathbb{Z})$. Two generators suffice:
\begin{equation}
    S\colon\; \tau \;\to\; -\frac{1}{\tau}\,,
    \qquad
    T\colon\; \tau \;\to\; \tau + 1\,,
    \label{eq:ST_generators}
\end{equation}
satisfying $S^2 = (ST)^3 = \mathbb{I}$. The transformation $S$ 
reflects through the unit circle while $T$ shifts $\tau$ by one 
unit along the real axis.

All physically inequivalent values of the modulus are contained 
in the fundamental domain of $\Gamma$,
\begin{equation}
    \mathcal{D} \;=\; 
    \left\{\,\tau \in \mathbb{H} \;\Big|\; 
    |\mathrm{Re}\,\tau| \leq \tfrac{1}{2}\,,
    \quad |\tau| \geq 1 \,\right\}\,,
    \label{eq:fund_domain}
\end{equation}
shown in Fig.~\ref{fig:tau_domain}. Its boundary consists of two 
vertical lines $\mathrm{Re}\,\tau = \pm\tfrac{1}{2}$ identified 
with each other under $T$, and the arc of the unit circle 
$|\tau|=1$ connecting $\omega = e^{2\pi i/3}$ to 
$\omega' = e^{i\pi/3}$, whose two halves are identified under $S$.
The domain $\mathcal{D}$ contains exactly two inequivalent points 
with non-trivial stabilisers~\cite{Serre1973}: $\tau = i$, 
fixed under $S$ with residual symmetry $\mathbb{Z}_2^S$, and 
$\tau = \omega$, fixed under $ST$ with residual symmetry 
$\mathbb{Z}_3^{ST}$.

The finite modular group entering the present model arises 
as the quotient $\Gamma_N \equiv \Gamma/\Gamma(N)$ evaluated 
at level $N=4$, yielding $\Gamma_4 \simeq S_4$. The 
principal congruence subgroup $\Gamma(N)$ is defined as
\begin{equation}
    \Gamma(N) \;=\; 
    \left\{\, 
    \begin{pmatrix} a & b \\ c & d \end{pmatrix} 
    \!\in\! \mathrm{SL}(2,\mathbb{Z}) \;\Bigg|\; 
    \begin{pmatrix} a & b \\ c & d \end{pmatrix} 
    \!\equiv\! 
    \begin{pmatrix} 1 & 0 \\ 0 & 1 \end{pmatrix} 
    \!\!\pmod{N} \right\}.
    \label{eq:congruence_sub}
\end{equation}
The group $S_4$ is of order 24 and its representation 
content consists of five irreducible representations: 
two singlets $\mathbf{1}$ and $\mathbf{1}'$, one 
doublet $\mathbf{2}$, and two triplets $\mathbf{3}$ 
and $\mathbf{3}'$~\cite{Novichkov:2018yse, 
Penedo:2018nmg}. The finite modular group 
$\Gamma_4 \simeq S_4$ is obtained from the 
infinite modular group $\bar{\Gamma} \equiv 
\mathrm{PSL}(2,\mathbb{Z})$ by imposing the 
additional level-4 constraint $T^4 = I$, which 
together with the relations $S^2 = (ST)^3 = I$ 
already satisfied in $\bar{\Gamma}$, truncates 
the infinite group to a finite group of 
order~24~\cite{Feruglio(2018)}.

A weight-$k$ modular form at level $N=4$ is defined 
as a holomorphic map $f:\mathbb{H}\to\mathbb{C}$ 
that responds to any $\gamma \in \Gamma(4)$ according to
\begin{equation}
    f(\gamma\tau) \;=\; (c\tau+d)^k\, f(\tau)\,, 
    \qquad \gamma \in \Gamma(4)\,.
\end{equation}
Following~\cite{Feruglio(2018), Novichkov:2018yse}, 
we impose modular invariance term by term in the 
Lagrangian. Matter fields are assigned to 
$\Gamma_4$ representations and carry zero modular 
weight, so they transform without an automorphy 
factor. The Yukawa couplings, by contrast, are 
identified with holomorphic modular forms of 
weight $k$ and therefore pick up a factor 
$(c\tau+d)^k$ under $\gamma \in \Gamma$; demanding 
that each Lagrangian term be invariant then fixes 
the total weight of every operator to zero.
Weight-$2$ modular forms at level $N=4$ span a 
five-dimensional complex vector space. Following Ref.~\cite{Penedo:2018nmg}, 
a basis of five generators $Y_1(\tau),\ldots,Y_5(\tau)$ 
can be chosen such that $(Y_1,\,Y_2)^T$ transforms as the 
doublet $\mathbf{2}$ of $S_4$ and $(Y_3,\,Y_4,\,Y_5)^T$ 
transforms as the triplet $\mathbf{3'}$. Higher-weight 
modular forms are constructed as homogeneous polynomials 
in $Y_1,\ldots,Y_5$; the relevant weight-4 combinations 
are listed in Table~\ref{tab:yukawa}. Restricting $\tau$ to $\mathcal{D}$ removes the redundancy 
from modular transformations: any two values related by 
$\gamma \in \Gamma$ yield identical 
observables~\cite{Novichkov:2018yse}. The explicit $q$-expansions of the weight-2 generators $Y_1(\tau),\ldots,Y_5(\tau)$, 
the decomposition of the weight-4 forms in Table~2 in terms of products of 
$Y_1,\ldots,Y_5$, and the relevant Clebsch--Gordan coefficients for $S_4$ 
tensor products are given in Refs.~\cite{Novichkov:2018yse,Penedo:2018nmg}.

\subsection{Particle Content and Symmetries}

The underlying symmetry of the framework consists of the SM gauge group $ SU(3)_C \times SU(2)_L \times U(1)_Y$, together with a discrete $\mathbb{Z}_2$ parity and modular invariance under the finite modular group $S_4$~\cite{Abhishek:2025ety}. Unlike conventional 
discrete flavour models, no flavon fields are 
introduced; instead, the Yukawa couplings are 
fixed by weight-$k$ holomorphic modular forms of 
$\tau$, each furnishing a representation of $S_4$, 
reducing the flavour freedom to the single complex 
parameter $\tau$. Radiative neutrino masses arise 
through the scotogenic loop~\cite{Ma:2006km}, while 
the baryon asymmetry is produced via flavoured 
resonant leptogenesis~\cite{Samanta:2019yeg}. 
Modular form conventions are taken from 
Ref.~\cite{Abhishek:2025ety}; field content and 
charge assignments appear in 
Tables~\ref{tab:matter} and~\ref{tab:yukawa}. Under $S_4$, the three lepton doublets $L$ transform 
as a triplet $\mathbf{3}$, whereas the right-handed 
fields $e_R$, $\mu_R$, $\tau_R$ are placed in 
$\mathbf{1}'$, $\mathbf{1}$, $\mathbf{1}'$. 
The charged-lepton mass matrix is then completely 
determined by the modular forms with no additional 
freedom.

Following the standard modular-invariance 
construction~\cite{Feruglio(2018)}, matter fields 
are given negative modular weights and Yukawa 
couplings are identified with modular forms of 
equal and opposite weight, so that each Lagrangian 
term carries total weight zero. The right-handed 
neutrino sector comprises two fields $N_R$ 
arranged as an $S_4$ doublet $\mathbf{2}$; with 
only two singlet fermions, $M_\nu$ is rank two 
and one neutrino remains massless at leading order. 
The $\mathbb{Z}_2$ parities are assigned as 
$N_R, \eta \to -1$ and all SM fields $\to +1$, 
simultaneously prohibiting a tree-level Dirac 
mass. After electroweak symmetry breaking (EWSB),
\begin{equation}
  \langle H \rangle = \frac{v}{\sqrt{2}}, \qquad \langle \eta \rangle = 0,
  \label{eq:vev}
\end{equation}
so that no neutrino mass is generated at the renormalisable tree
level. The $\mathbb{Z}_2$ symmetry assigns odd parity to both 
$N_R$ and $\eta$ while all SM fields carry even 
parity. Consequently, the tree-level Dirac mass 
term $\bar{L}\tilde{H}N_R$ is forbidden since 
$H$ is $\mathbb{Z}_2$-even, whereas the scotogenic Yukawa 
coupling $\bar{L}\tilde{\eta}N_R$ is permitted 
since both $\eta$ and $N_R$ are odd. The $\mathbb{Z}_2$ symmetry simultaneously stabilises the lightest 
odd-sector particle, providing a built-in DM candidate.
\begin{table}[t]
\centering
\caption{Charge assignments and
Modular weights $k_I$ of the matter and scalar fields.}
\label{tab:matter}
\begin{tabular}{lcccccccc}
\hline
Field   & $L$ & $l_{R_1}$ & $l_{R_2}$ & $l_{R_3}$ & $N_R$ & $H$ & $\eta$ \\
\hline
$SU(2)_L$ & 2 & 1 & 1 & 1 & 1 & 2 & 2 \\
$U(1)_Y$  & $-\tfrac{1}{2}$ & $-1$ & $-1$ & $-1$ & 0 & $\tfrac{1}{2}$ & $\tfrac{1}{2}$ \\
$S_4$     & $\mathbf{3}$ & $\mathbf{1}'$ & $\mathbf{1}$ & $\mathbf{1}'$ & $\mathbf{2}$ & $\mathbf{1}$ & $\mathbf{1}$ \\
$\mathbb{Z}_2$     & $+$ & $+$ & $+$ & $+$ & $-$ & $+$ & $-$ \\
$k_I$     & $-2$ & $0$ & $-2$ & $-2$ & $-2$ & $0$ & $0$ \\
\hline
\end{tabular}
\end{table}

\begin{table}[t]
\centering
\caption{Modular Yukawa couplings
with modular weight $k_I$ gives the form weight.}
\label{tab:yukawa}
\begin{tabular}{lcccccc}
\hline
 & $Y^{(2)}_{\mathbf{3}}$ & $Y^{(2)}_{\mathbf{3}'}$ & $Y^{(4)}_{\mathbf{1}}$ & $Y^{(4)}_{\mathbf{3}}$ & $Y^{(4)}_{\mathbf{3}'}$ & $Y^{(4)}_{\mathbf{2}}$ \\
\hline
$S_4$ & $\mathbf{3}$ & $\mathbf{3}'$ & $\mathbf{1}$ & $\mathbf{3}$ & $\mathbf{3}'$ & $\mathbf{2}$ \\
$k_I$ & 2 & 2 & 4 & 4 & 4 & 4 \\
\hline
\end{tabular}
\end{table}
\subsection{Yukawa Sector}
\label{sec:2.1}

All Yukawa interactions are assembled from modular forms so as to
preserve modular invariance of the Lagrangian. The charged-lepton
sector is governed by
\begin{equation}
  \mathcal{L}_\ell
  = \alpha\,(\bar{L}e_R)_{\mathbf{3}'}H\,Y^{(2)}_{\mathbf{3}'}
  + \beta\,(\bar{L}\mu_R)_{\mathbf{3}}H\,Y^{(4)}_{\mathbf{3}}
  + \gamma\,(\bar{L}\tau_R)_{\mathbf{3}'}H\,Y^{(4)}_{\mathbf{3}'}
  + \mathrm{h.c.},
  \label{eq:lag_cl}
\end{equation}
where $\alpha$, $\beta$, $\gamma$ are dimensionless $\mathcal{O}(1)$
coefficients and $(\cdots)_r$ denotes contraction into the
irreducible $S_4$ representation $r$. EWSB yields the 
charged-lepton mass matrix
\begin{equation}
  M_\ell = \frac{v}{\sqrt{2}}
  \begin{pmatrix}
    \alpha Y_3     & -2\beta Y_2 Y_3                          & 2\gamma Y_1 Y_3 \\
    \alpha Y_5     & \beta(\sqrt{3}Y_1 Y_4 + Y_2 Y_5)        & \gamma(\sqrt{3}Y_2 Y_4 - Y_1 Y_5) \\
    \alpha Y_4     & \beta(\sqrt{3}Y_1 Y_5 + Y_2 Y_4)        & \gamma(\sqrt{3}Y_2 Y_5 - Y_1 Y_4)
  \end{pmatrix}.
  \label{eq:Mell}
\end{equation}

The Dirac neutrino Yukawa interactions take the form
\begin{equation}
  \mathcal{L}_\nu
  = \alpha_D\,(\bar{L}N_R)_{\mathbf{3}}\,\tilde{\eta}\,\tilde{Y}^{(4)}_{\mathbf{3}}
  + \beta_D\,(\bar{L}N_R)_{\mathbf{3}'}\,\tilde{\eta}\,\tilde{Y}^{(4)}_{\mathbf{3}'}
  + \mathrm{h.c.},
  \label{eq:lag_nu}
\end{equation}
where $\tilde{\eta} = i\sigma_2\eta^*$. The resulting $3\times 2$
Dirac mass matrix is
\begin{equation}
  M_D = \frac{v}{\sqrt{2}}
\begin{pmatrix}
-2\alpha_D p & -2\beta_D q \\
-\alpha_D\!\left(\frac{\sqrt{3}}{2}t + \frac{1}{2}s\right) 
+ \beta_D\!\left(\frac{3}{2}s - \frac{\sqrt{3}}{2}t\right) &
\alpha_D\!\left(\frac{3}{2}u + \frac{\sqrt{3}}{2}r\right) 
+ \beta_D\!\left(\frac{\sqrt{3}}{2}r - \frac{1}{2}u\right) \\
-\alpha_D\!\left(\frac{\sqrt{3}}{2}u + \frac{1}{2}r\right) 
+ \beta_D\!\left(\frac{3}{2}r - \frac{\sqrt{3}}{2}u\right) &
\alpha_D\!\left(\frac{3}{2}t + \frac{\sqrt{3}}{2}s\right) 
+ \beta_D\!\left(\frac{\sqrt{3}}{2}s - \frac{1}{2}t\right)
\end{pmatrix},
  \label{eq:MD}
\end{equation}
with the shorthand
\begin{equation}
  p=Y_2 Y_3,\quad q=Y_1 Y_3,\quad r=Y_2 Y_4,\quad
  s=Y_2 Y_5,\quad t=Y_1 Y_4,\quad u=Y_1 Y_5.
  \label{eq:shorthand}
\end{equation}

Using two-component Weyl spinors throughout, 
the $\Gamma(4)$-invariant Majorana mass 
Lagrangian for $N_R$ is
\begin{equation}
  \mathcal{L}_R
  = \tfrac{1}{2}M Y^{(4)}_{\mathbf{1}}\,(N_R^c N_R)
  + \tfrac{1}{2}M_\epsilon
    \!\left(N_R^c\cdot Y^{(4)}_{\mathbf{2}}\otimes N_R\right)_{\mathbf{1}}
  + \mathrm{h.c.},
  \label{eq:lag_R}
\end{equation}
which gives the $2\times 2$ mass matrix
\begin{equation}
  M_R =
  \begin{pmatrix}
    M(Y_1^2+Y_2^2) - M_\epsilon(Y_2^2-Y_1^2) & M_\epsilon Y_1 Y_2 \\
    M_\epsilon Y_1 Y_2 & M(Y_1^2+Y_2^2) + M_\epsilon(Y_2^2-Y_1^2)
  \end{pmatrix}.
  \label{eq:MR}
\end{equation}
In the regime $M_\epsilon \ll M$, the off-diagonal terms of $M_R$ are subdominant relative to the diagonal entries. Consequently, the heavy neutrino mass eigenstates become nearly degenerate, giving rise to an intrinsically quasi-degenerate spectrum. 
The parameter $M_\epsilon$ controls the departure from exact
degeneracy in the heavy Majorana neutrino sector. In the limit
$M_\epsilon \to 0$, the Majorana mass matrix becomes
proportional to the identity matrix,
\begin{equation}
M_R \to
M\left(Y_1^2+Y_2^2\right)\mathbb{I}_2,
\end{equation}
leading to two exactly degenerate heavy neutrino masses,
\begin{equation}
M_1=M_2=
M\left(Y_1^2+Y_2^2\right).
\end{equation}
For non-zero but sufficiently small $M_\epsilon$, the degeneracy
is lifted and the physical masses become
\begin{equation}
M_{1,2}
=
M\left(Y_1^2+Y_2^2\right)
\mp
M_\epsilon
\sqrt{
Y_1^4-Y_1^2Y_2^2+Y_2^4
}.
\label{eq:MR_eigenvalues}
\end{equation}
The corresponding mass splitting is therefore
\begin{equation}
\Delta M \equiv M_2-M_1
=
2M_\epsilon
\sqrt{
Y_1^4-Y_1^2Y_2^2+Y_2^4
}.
\label{eq:DM_exact}
\end{equation}
Thus, for $M_\epsilon \ll M$, the splitting satisfies
$\Delta M \ll M$, yielding a naturally quasi-degenerate heavy
neutrino spectrum suitable for resonant leptogenesis. The near-degenerate heavy neutrino spectrum therefore follows
directly from the structure of the Majorana mass matrix and is
controlled by the single parameter $M_\epsilon$. This provides
the required condition for resonant enhancement of the CP
asymmetry in leptogenesis without introducing additional
ad hoc fine tuning of the heavy neutrino masses.

\subsection{Scalar Potential and Mass Spectrum}
\label{sec:2.2}

The scalar sector consists of $H\sim 
(\mathbf{2},\tfrac{1}{2})$ and 
$\eta\sim(\mathbf{2},\tfrac{1}{2})$ under 
$SU(2)_L\times U(1)_Y$, with $\eta$ odd under 
$\mathbb{Z}_2$. The full renormalisable potential 
consistent with all symmetries reads
\begin{align}
  V(H,\eta)
  &= \mu_1^2 H^\dagger H + \mu_2^2 \eta^\dagger\eta
  + \lambda_1(H^\dagger H)^2 + \lambda_2(\eta^\dagger\eta)^2
  \nonumber\\
  &\quad
  + \lambda_3(H^\dagger H)(\eta^\dagger\eta)
  + \lambda_4(H^\dagger\eta)(\eta^\dagger H)
  + \frac{\lambda_5}{2}\!\left[(H^\dagger\eta)^2 + \mathrm{h.c.}\right].
  \label{eq:potential}
\end{align}
Expanding around the electroweak vacuum
$\langle H\rangle = v/\sqrt{2}$, $\langle\eta\rangle = 0$,
the minimisation condition requires $\mu_1^2 = -\lambda_1 v^2$.
The physical mass eigenvalues are
\begin{equation}
  m_h^2 = 2\lambda_1 v^2, \qquad
  m_{\eta^\pm}^2 = \mu_2^2 + \tfrac{v^2}{2}\lambda_3,
  \label{eq:masses1}
\end{equation}
\begin{equation}
  m_{\eta_R}^2 = \mu_2^2 + \tfrac{v^2}{2}(\lambda_3+\lambda_4+\lambda_5),
  \qquad
  m_{\eta_I}^2 = \mu_2^2 + \tfrac{v^2}{2}(\lambda_3+\lambda_4-\lambda_5).
  \label{eq:masses2}
\end{equation}
The neutral scalar mass splitting is therefore
\begin{equation}
  m_{\eta_R}^2 - m_{\eta_I}^2 = \lambda_5 v^2,
  \label{eq:splitting}
\end{equation}
controlled entirely by $\lambda_5$. This splitting drives the
one-loop radiative neutrino mass: the two neutral components
$\eta_R$ and $\eta_I$ circulate in the loop with opposite-sign
contributions, and the difference is proportional to $\lambda_5$.
In the exact limit $\lambda_5\to 0$ the two components become
mass-degenerate and their contributions cancel, leaving a vanishing
neutrino mass. Non-zero $\lambda_5$ is therefore a necessary
condition for generating neutrino masses within this framework.

The following tree-level conditions on the 
quartic couplings are necessary for vacuum 
stability and perturbative 
unitarity~\cite{Ma:2006km}:
\begin{equation}
  \lambda_1>0,\quad \lambda_2>0,\quad
  \lambda_3+\sqrt{\lambda_1\lambda_2}>0,\quad
  \lambda_3+\lambda_4-|\lambda_5|+\sqrt{\lambda_1\lambda_2}>0.
  \label{eq:stability}
\end{equation}
A further constraint comes from demanding that $\eta$ does not
develop a VEV. After EWSB, the effective squared mass of the inert doublet receives a
shift $\frac{v^2}{2}\lambda_3$ from the $\lambda_3$ interaction.
The condition $\langle\eta\rangle=0$ is equivalent to
\begin{equation}
  \mu_2^2 + \tfrac{v^2}{2}\lambda_3 > 0,
  \label{eq:inert}
\end{equation}
which ensures that the $\mathbb{Z}_2$ symmetry remains unbroken at the
electroweak minimum. We note that eq.~\eqref{eq:inert} is a 
tree-level stability condition; for the 
perturbative values of $\lambda_{3,4,5}$ 
considered in this work, loop corrections 
to $m_\eta^2$ are suppressed by 
$1/16\pi^2$ and do not destabilise the 
$Z_2$-preserving vacuum~\cite{Ma:2006km}. With this condition satisfied, the lightest
$\mathbb{Z}_2$-odd state is absolutely stable. 

\section{Neutrino Phenomenology}
\label{sec:3}
\begin{table}[t]
\centering
\caption{Neutrino oscillation parameters: best-fit values and $3\sigma$ allowed ranges from the NuFIT~5.2 (2024) global analysis assuming normal mass ordering~\cite{Esteban:2024vvu}.}
\label{tab:nufit}
\begin{tabular}{lcc}
\hline
Parameter & Best-fit $\pm\,1\sigma$ & $3\sigma$ range \\
\hline
$\sin^2\theta_{12}$ & $0.307^{+0.012}_{-0.011}$ & $0.275-0.345$ \\
$\sin^2\theta_{23}$ & $0.561^{+0.012}_{-0.015}$ & $0.430-0.596$ \\
$\sin^2\theta_{13}$ & $0.02195^{+0.00054}_{-0.00058}$ & $0.02023-0.02376$ \\
$\Delta m^2_{21}\;[10^{-5}\,\mathrm{eV}^2]$ & $7.49^{+0.19}_{-0.19}$ & $6.92-8.05$ \\
$\Delta m^2_{31}\;[10^{-3}\,\mathrm{eV}^2]$ & $2.534^{+0.025}_{-0.023}$ & $2.463-2.606$ \\
\hline
\end{tabular}
\end{table}

We now analyse the neutrino phenomenology of the 
model. Since neutrino masses arise at one loop, 
their structure is entirely fixed by $\tau$ and 
the Yukawa couplings, strongly correlating all 
leptonic observables with the modular parameter 
space. Consequently, all neutrino observables are
strongly correlated with the modular parameter space, making the
framework highly predictive.

The heavy Majorana mass matrix $M_R$ is diagonalised by a unitary matrix $U_R$ according to
\begin{equation}
U_R^T M_R U_R = \mathrm{diag}(M_1,M_2),
\end{equation}
where $M_i$ are the physical heavy neutrino masses. Working in the heavy-neutrino mass basis requires a corresponding rotation of the Dirac Yukawa matrix,
\begin{equation}
\tilde{Y}_\nu = Y_\nu U_R,
\end{equation}
so that the Yukawa interactions and heavy neutrino propagators are expressed in terms of the physical mass eigenstates.

Following EWSB, the mass splitting between the neutral components of the inert scalar doublet induced by the scalar potential enables the radiative generation of light neutrino masses. Consequently, the interaction
\begin{equation}
  \mathcal{L} \supset (Y_\nu)_{\alpha i}\,\bar{L}_\alpha\,\tilde{\eta}\,N_i
  + \tfrac{1}{2}N_i^T M_{R\,ij} N_j + \mathrm{h.c.}
\end{equation}
gives rise to a Majorana mass matrix for the light neutrinos at the one-loop level. Working throughout in the heavy-mass eigenstate
basis, this matrix takes the form
\begin{equation}
  (M_\nu)_{\alpha\beta}
  = \sum_{i=1}^{2}(\tilde{Y}_\nu)_{\alpha i}\,(\tilde{Y}_\nu)_{\beta i}\,\Lambda_i,
  \label{eq:loopmass}
\end{equation}
where the loop integral evaluates to~\cite{Ma:2006km}
\begin{equation}
  \Lambda_i = \frac{M_i}{16\pi^2}
  \left[
    \frac{m_{\eta_R}^2}{m_{\eta_R}^2 - M_i^2}
    \ln\!\left(\frac{m_{\eta_R}^2}{M_i^2}\right)
    -
    \frac{m_{\eta_I}^2}{m_{\eta_I}^2 - M_i^2}
    \ln\!\left(\frac{m_{\eta_I}^2}{M_i^2}\right)
  \right].
  \label{eq:Lambda}
\end{equation}

The two terms in Eq.~\eqref{eq:Lambda} arise from the CP-even and
CP-odd neutral inert scalars circulating in the loop. Their
difference is controlled by $\lambda_5$ through the mass splitting
$m_{\eta_R}^2 - m_{\eta_I}^2 = \lambda_5 v^2$; in the limit
$\lambda_5\to 0$ the loop function $\Lambda_i$ vanishes identically,
recovering the tree-level result of a massless neutrino sector. A
finite $\lambda_5$ is therefore not merely a technical assumption
but a physical requirement for the mechanism to operate.

Since there are only two RHNs in the model, the lightest neutrino eigenstate is massless at leading order because the matrix $M_\nu$ in Eq.~\eqref{eq:loopmass} has rank two. The massless state is represented by $m_1 = 0$ in the numerical analysis, which is performed under the assumption of normal mass ordering (NO).

The charged-lepton mass matrix is brought to diagonal form by a
unitary transformation,
\begin{equation}
  U_\ell^\dagger M_\ell M_\ell^\dagger U_\ell
  = \mathrm{diag}(m_e^2,\,m_\mu^2,\,m_\tau^2),
  \label{eq:Melldiag}
\end{equation}
while the Majorana neutrino mass matrix satisfies
\begin{equation}
  U_\nu^T M_\nu U_\nu = \mathrm{diag}(m_1,\,m_2,\,m_3).
  \label{eq:Mnudiag}
\end{equation}
The PMNS matrix
\begin{equation}
U_{\rm PMNS} = U_\ell^\dagger U_\nu,
\label{eq:UPMNS}
\end{equation}
is parametrised following the PDG 
convention~\cite{ParticleDataGroup:2024cfk} 
in terms of three mixing angles 
$(\theta_{12},\theta_{23},\theta_{13})$, one 
Dirac phase $\delta$, and two Majorana phases. 
For normal ordering,
\begin{equation}
\Delta m^2_{21} = m_2^2 - m_1^2, \qquad
\Delta m^2_{31} = m_3^2 - m_1^2.
\label{eq:masssq}
\end{equation}

The three mixing angles are read off from the PMNS matrix elements as
\begin{equation}
  \sin^2\theta_{13} = |U_{e3}|^2,\qquad
  \sin^2\theta_{12} = \frac{|U_{e2}|^2}{1-|U_{e3}|^2},\qquad
  \sin^2\theta_{23} = \frac{|U_{\mu 3}|^2}{1-|U_{e3}|^2}.
  \label{eq:angles}
\end{equation}
The Dirac CP phase is extracted through the Jarlskog invariant~\cite{Jarlskog:1985cw, Bilenky:1987ty,Krastev:1988yu},
\begin{equation}
J_{CP} =
\mathrm{Im}
\left[
U_{e1} U_{\mu2} U_{e2}^* U_{\mu1}^*
\right],
\end{equation}
which satisfies
\begin{equation}
J_{CP} =
\frac{1}{8}
\sin 2\theta_{12}
\sin 2\theta_{23}
\sin 2\theta_{13}
\cos\theta_{13}
\sin\delta .
\end{equation}

Two further observables of direct experimental relevance are
computed. The effective Majorana mass governing the rate of
neutrinoless double beta decay ($0\nu\beta\beta$) is~\cite{Abhishek:2026hex}
\begin{equation}
  m_{\beta\beta}
  = \left|\sum_{i=1}^{3} U_{ei}^2\,m_i\right|,
  \label{eq:mbb}
\end{equation}
and the sum of neutrino masses $\sum_i m_i$ is constrained by
cosmological observations~\cite{Planck:2018vyg}.

\begin{table}[t]
\centering
\caption{Scan ranges for the free parameters of the model.
Only solutions reproducing the neutrino oscillation
data within the $3\sigma$ intervals of Table~\ref{tab:nufit}
are retained.}
\label{tab:scan}
\begin{tabular}{lc}
\hline
Parameter & Scan range \\
\hline
$\mathrm{Re}(\tau)$ & $[-0.5,\;0.5]$ \\
$\mathrm{Im}(\tau)$ & $[0.9,\;2]$ \\
$\alpha$ & $[0.5,\;1.6]$ \\
$\beta$  & $[0.3,\;1.1]$ \\
$\gamma$ & $[1.0,\;1.6]$ \\
$\alpha_D$ & $[10^{-6},\;5\times10^{-3}]$ \\
$\beta_D$  & $[10^{-5},\;5\times10^{-3}]$ \\
$M\;[\mathrm{GeV}]$ & $[10^4,\;5\times10^5]$ \\
$M_\epsilon\;[\mathrm{GeV}]$ & $[10^{-2},\;1]$ \\
$m_\eta\;[\mathrm{GeV}]$ & $[1,\;10^3]$ \\
$\lambda_5$ & $[10^{-5},\;10^{-2}]$ \\
\hline
\end{tabular}
\label{tab:scanranges}
\end{table}

\begin{figure}[t]
    \centering
    \includegraphics[width=0.55\textwidth]{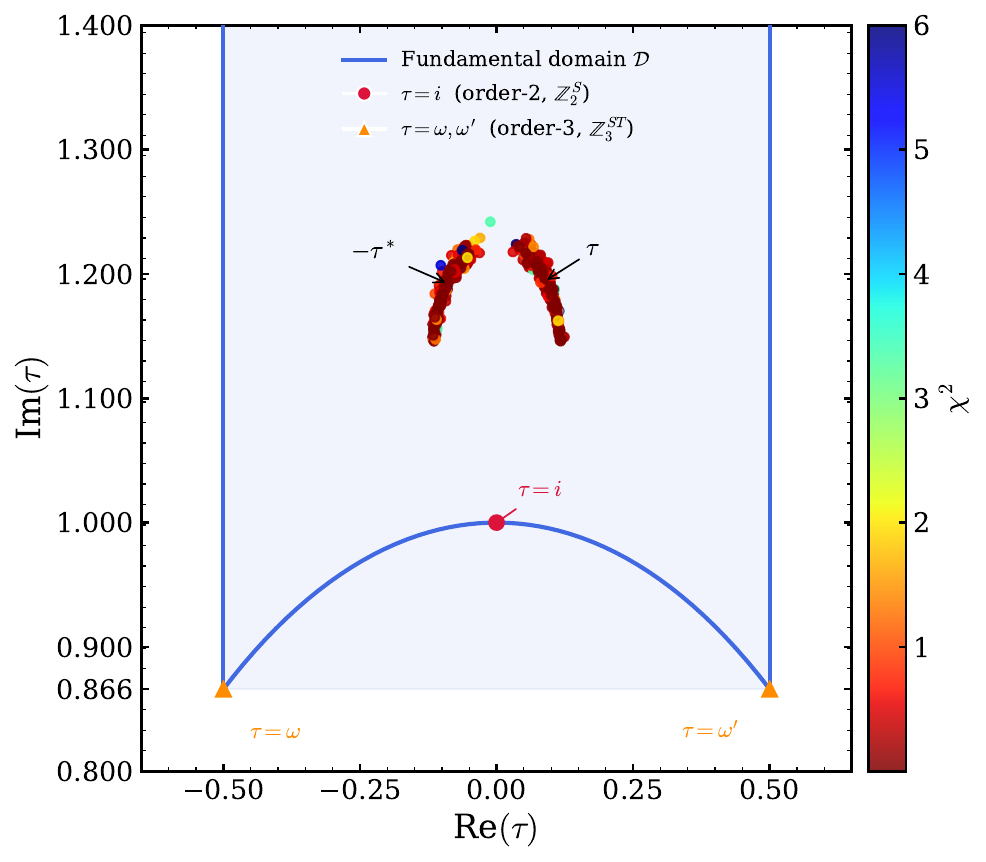}
    \caption{Distribution of viable parameter points in the 
    fundamental domain $\mathcal{D}$ of the modular group, 
    coloured by $\chi^2$. The blue boundary is the standard 
    domain $|\mathrm{Re}\,\tau|\leq\tfrac{1}{2}$, $|\tau|\geq 1$. 
    The two clusters at $\mathrm{Re}(\tau)\approx\pm 0.14$ are 
    related by the conjugation $\tau\to -\tau^*$, predicting 
    opposite CP phases with identical mixing angles. 
    Fixed points $\tau=i$ ($\mathbb{Z}_2^S$) and 
    $\tau=\omega$ ($\mathbb{Z}_3^{ST}$) are marked but 
    lie away from the viable region.}
    \label{fig:tau_domain}
\end{figure}

The scan explores a broader region of modular parameter space,
including modular images related by $T$-transformations. After mapping the
viable solutions into the fundamental domain $\mathcal{D}$, all accepted points
cluster nearby $\mathrm{Re}(\tau)\simeq \pm 0.14$ and
$\mathrm{Im}(\tau)\simeq 1.15\text{--}1.23$, as shown in
Fig.~\ref{fig:tau_domain}. The two allowed regions of parameter space are connected through the transformation
$\tau \rightarrow -\tau^{*}$, under which the neutrino mixing angles remain invariant, whereas the CP-violating phases change sign~\cite{Novichkov:2018yse}. Notably, the
viable solutions lie away from the special fixed points $\tau=i$ and
$\tau=\omega$, indicating the absence of residual modular symmetries in the
phenomenologically allowed region.

\begin{figure}[t]
\centering
\includegraphics[width=0.49\textwidth]{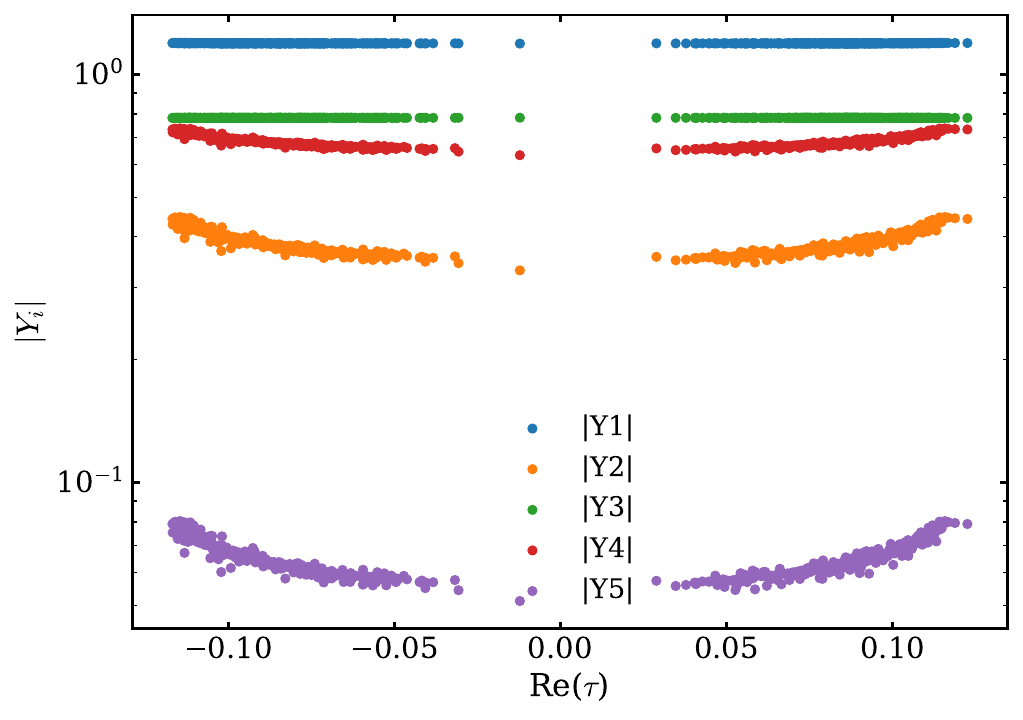}
\includegraphics[width=0.49\textwidth]{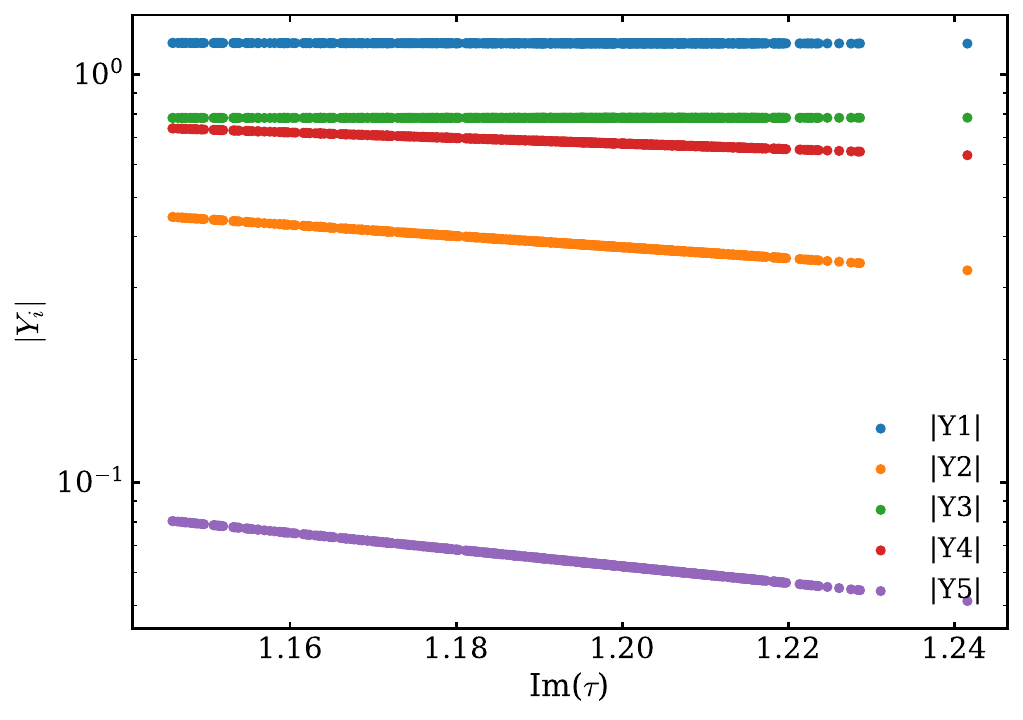}
\caption{
Magnitude of the modular Yukawa couplings $|Y_i|$ as a function of the modulus $\tau$.
The left panel shows the dependence on $\mathrm{Re}(\tau)$ while the right panel shows the dependence on $\mathrm{Im}(\tau)$.
}
\label{fig:Ytau}
\end{figure}

The dependence of the five independent modular Yukawa couplings
$|Y_i|$ on $\mathrm{Re}(\tau)$ and $\mathrm{Im}(\tau)$ is
illustrated in Fig.~\ref{fig:Ytau}, demonstrating the rich and
non-trivial flavour structure encoded by the modulus. The charged lepton Yukawa coefficients
$\alpha$, $\beta$, and $\gamma$ are taken to be $\mathcal{O}(1)$
parameters, consistent with the modular flavor framework. The Dirac
neutrino Yukawa couplings $\alpha_D$ and $\beta_D$ are scanned over
smaller values in order to reproduce the correct scale of radiatively
generated neutrino masses in the scotogenic mechanism. The heavy
RHN mass scale $M$ and the splitting parameter $M_e$
are varied such that the resulting mass spectrum allows for
quasi-degenerate heavy neutrinos, which is required for resonant
leptogenesis. The inert scalar doublet is characterized by a mass scale $m_\eta$ in the TeV range. A non-zero value of the coupling $\lambda_5$ lifts the degeneracy between its neutral CP-even and CP-odd components, providing the necessary ingredient for the radiative generation of light neutrino masses at the one-loop level.

For each parameter point, the light neutrino mass matrix is evaluated using Eq.~(\ref{eq:loopmass}) and subsequently diagonalized to extract the neutrino mass eigenvalues and leptonic mixing parameters. The resulting viable parameter space yields definite predictions for the neutrino mass spectrum, leptonic mixing parameters, and CP-violating phases in the modular $S_4$ scotogenic model.

\begin{figure}[t]
\centering
\begin{minipage}{0.95\textwidth}
\centering
\includegraphics[width=0.49\textwidth]{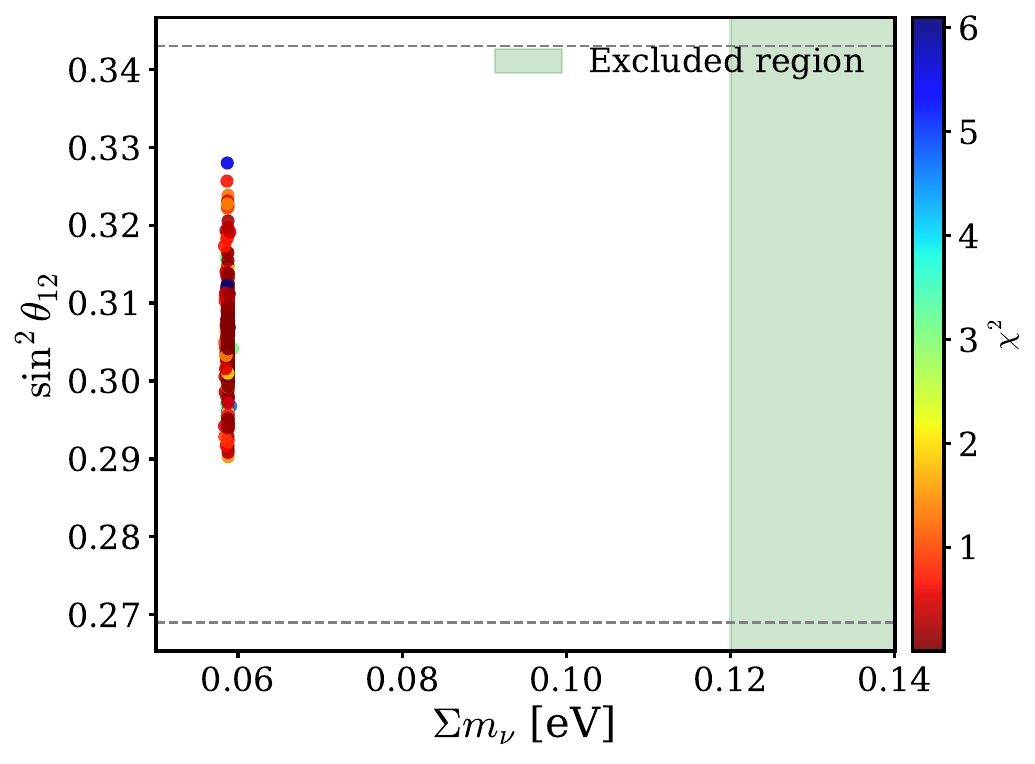}
\includegraphics[width=0.49\textwidth]{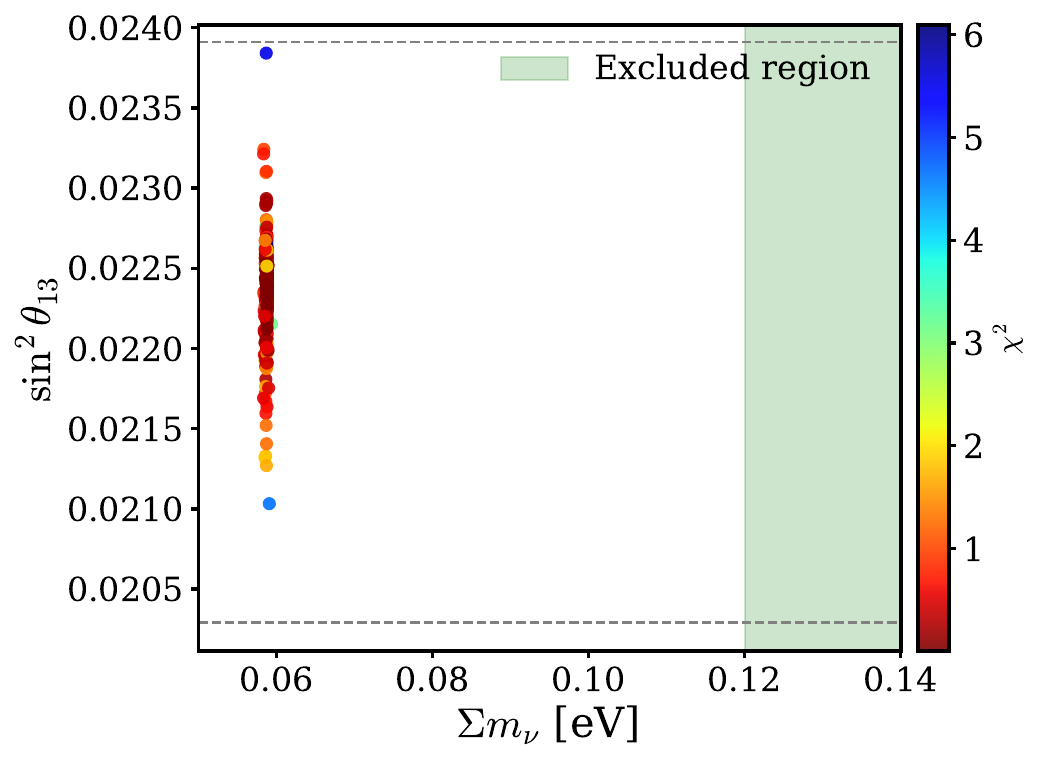}
\end{minipage}

\vspace{0.3cm}

\begin{minipage}{0.95\textwidth}
\centering
\includegraphics[width=0.49\textwidth]{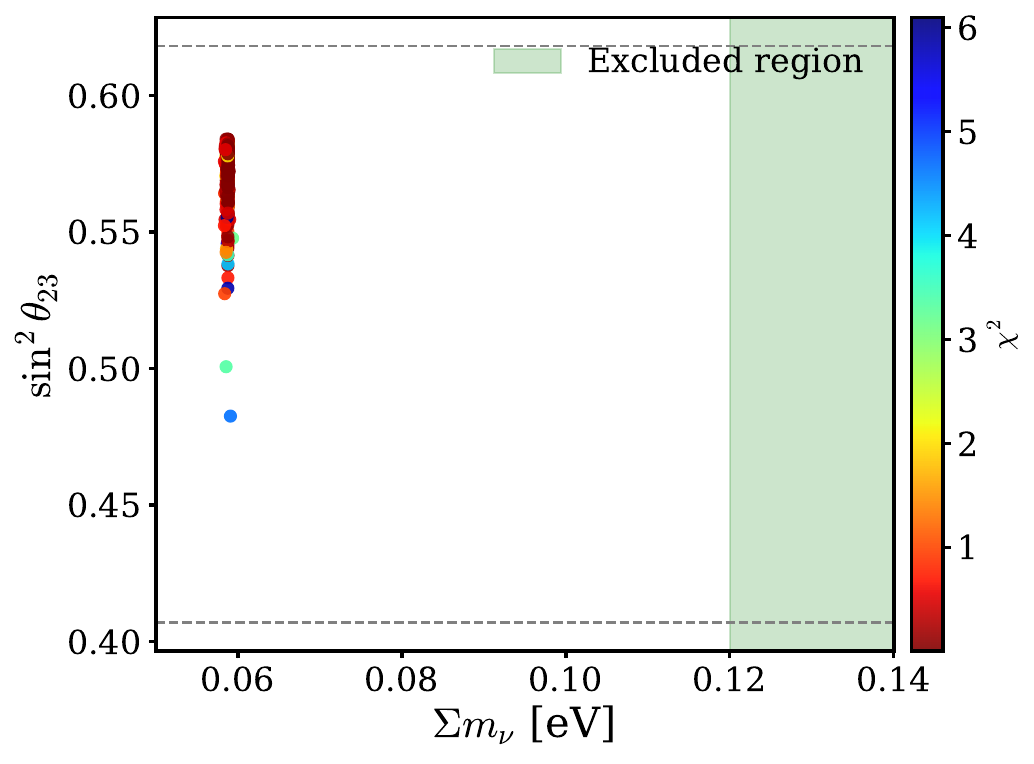}
\includegraphics[width=0.49\textwidth]{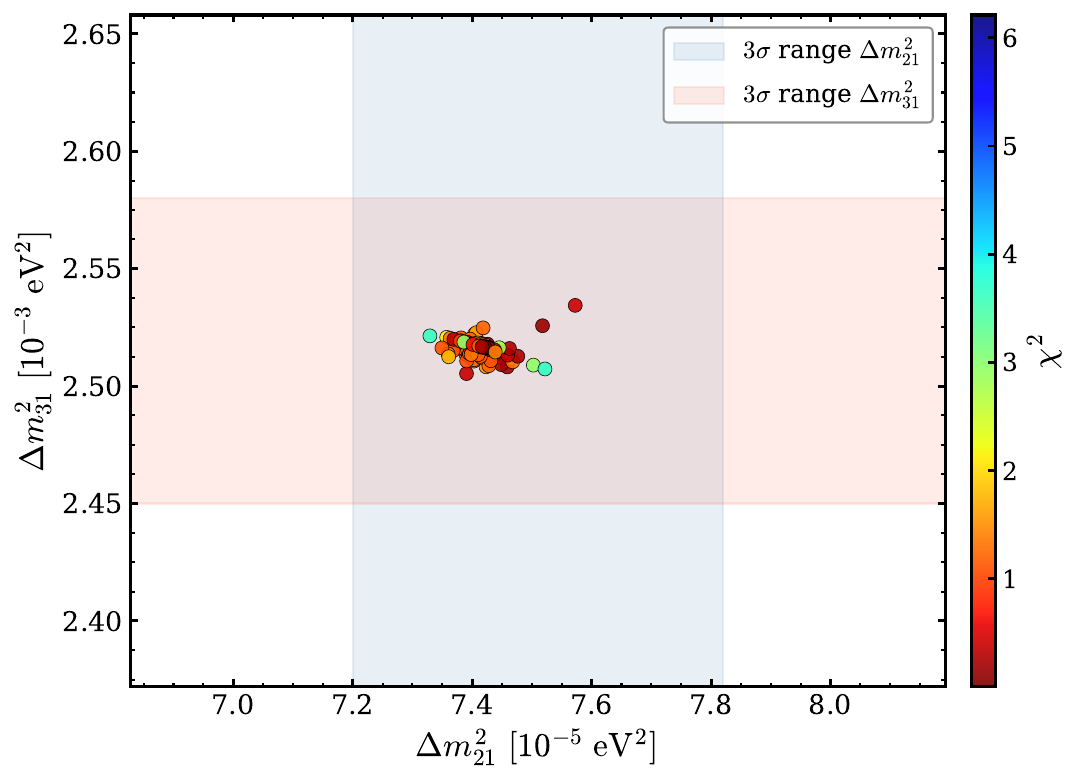}
\end{minipage}

\caption{Correlations of the neutrino mass sum 
$\sum m_\nu$ with the mixing parameters 
$\sin^2\theta_{12}$, $\sin^2\theta_{23}$, and 
$\sin^2\theta_{13}$. Dashed lines mark the NuFIT~5.2 
$3\sigma$ boundaries~\cite{Esteban:2024vvu} and the 
shaded band the cosmological bound 
$\sum m_\nu < 0.12\,$eV~\cite{Planck:2018vyg}. 
Bottom-right: correlation between $\Delta m^2_{21}$ 
and $\Delta m^2_{31}$; blue and red bands indicate 
their respective $3\sigma$ ranges.}
\label{fig:mixing_sum}
\end{figure}

Figure~\ref{fig:mixing_sum} shows the dependence of the total neutrino mass sum, $\sum m_\nu$, on the three leptonic mixing angles. The dashed horizontal lines denote the corresponding $3\sigma$ allowed intervals from the NuFIT global analysis, while the shaded region represents the cosmological constraint $\sum m_\nu < 0.12\,\mathrm{eV}$~\cite{Planck:2018vyg}.
The viable parameter points cluster in well-defined regions,
reflecting the strong constraints imposed by modular invariance.
The predicted ranges for the mixing angles are
\begin{equation}
0.290 \lesssim \sin^2\theta_{12} \lesssim 0.329 ,
\end{equation}
while the atmospheric mixing angle lies within
\begin{equation}
0.480 \lesssim \sin^2\theta_{23} \lesssim 0.584 .
\end{equation}
The model correctly accommodates the measured value of $\theta_{13}$, as evidenced by the reproduction of the reactor mixing angle in the interval
\begin{equation} 0.0210 \lesssim \sin^2\theta_{13} \lesssim 0.0237.
\end{equation}

Current global fits are also consistent with the neutrino mass-squared differences obtained from the numerical scan. In particular, the solar mass-squared difference is found to lie within
\begin{equation}
7.32\times10^{-5} \lesssim \Delta m_{21}^2 \lesssim 7.57\times10^{-5}\,\mathrm{eV}^2,
\end{equation}
while the atmospheric mass-squared difference falls in the range
\begin{equation}
2.50\times10^{-3} \lesssim \Delta m_{31}^2 \lesssim 2.53\times10^{-3}\,\mathrm{eV}^2.
\end{equation}

\begin{figure}[t]
\centering
\includegraphics[width=0.51\textwidth]{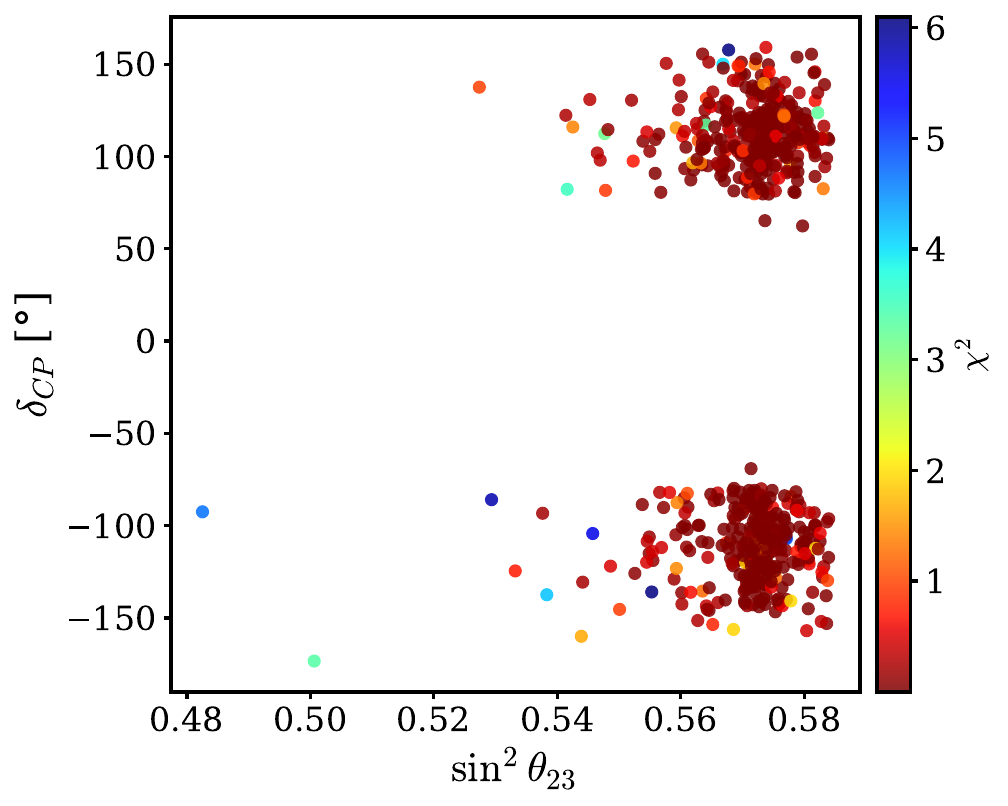}
\includegraphics[width=0.48\textwidth]{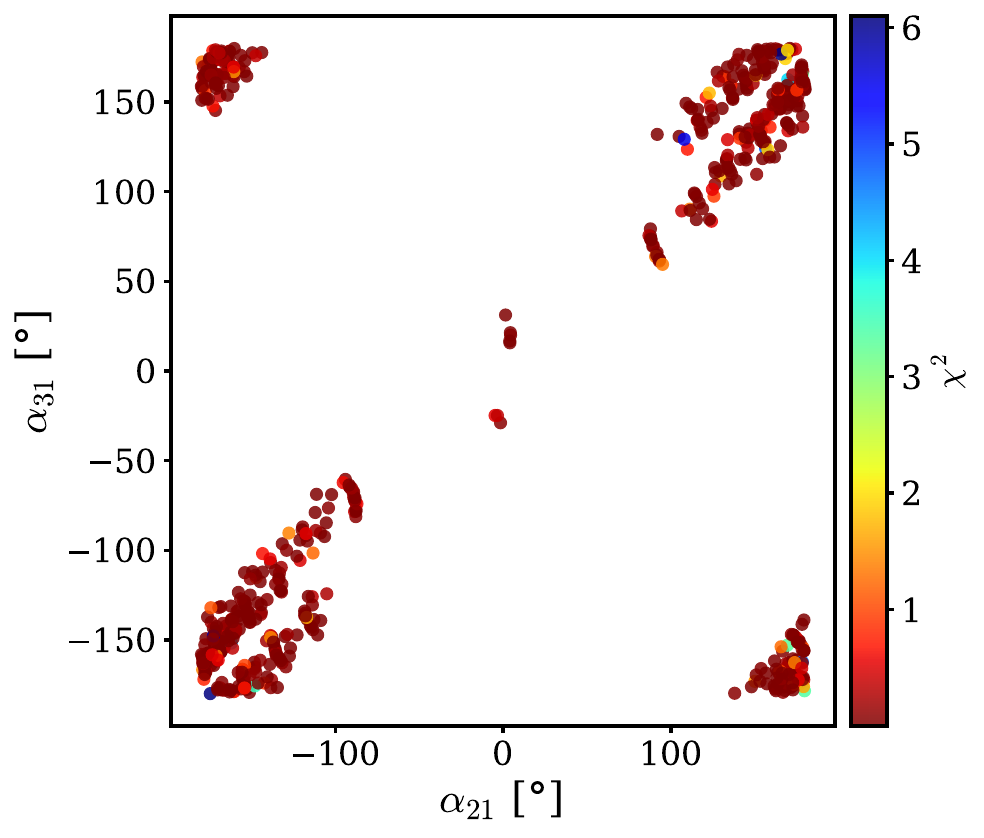}
\caption{
The behavior of leptonic CP phases predicted by the model is presented.
The left plot depicts how the Dirac phase $\delta_{CP}$ varies with the atmospheric mixing parameter $\sin^2\theta_{23}$, while the right plot highlights the interplay between the Majorana phases $\alpha_{21}$ and $\alpha_{31}$.
}
\label{fig:CP_phase}
\end{figure}

Figure~\ref{fig:CP_phase} illustrates the predicted correlations among the leptonic CP-violating phases. The left panel displays the correlation between the atmospheric mixing parameter $\sin^2\theta_{23}$ and the Dirac CP-violating phase $\delta_{\rm CP}$. The viable solutions are distributed across two broad regions
spanning $\delta_{CP} \simeq \pm(50^\circ\text{--}150^\circ)$,
with the densest concentration near
$\delta_{CP} \simeq \pm(90^\circ\text{--}150^\circ)$, reflecting the
underlying modular symmetry and the conjugation relation
$\tau \to -\tau^*$. The atmospheric mixing angle spans
$\sin^2\theta_{23} \simeq 0.48\text{--}0.58$,
with the bulk of solutions concentrated above $0.52$.

\begin{figure}[htbp]
\centering
\includegraphics[width=0.60\textwidth]{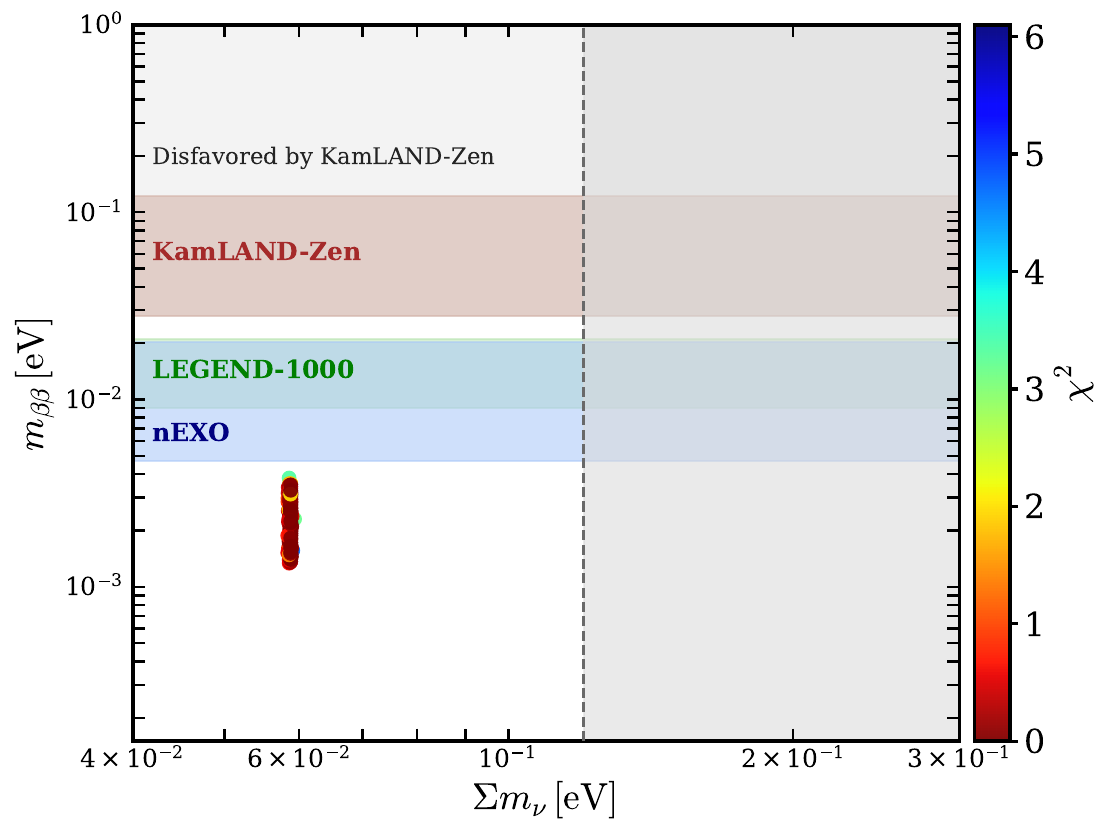}

\caption{Predicted $m_{\beta\beta}$ as a function 
of $\sum m_\nu$, coloured by $\chi^2$. Shaded 
bands indicate the KamLAND-Zen 
exclusion~\cite{KamLANDZen2016} and the 
prospective reaches of 
LEGEND-1000~\cite{LEGEND2021} and 
nEXO~\cite{nEXO2018}.
}
\label{fig:mbb_m1}
\end{figure}

The correlation between the Majorana phases $\alpha_{21}$ and $\alpha_{31}$ is seen in the right panel. The allowed parameter points
occupy highly localized regions in the phase plane, indicating strong
constraints imposed by the modular structure of the model. The non-trivial flavor structure of the neutrino mass matrix is reflected in the presence of many symmetric branches, which comes from the discrete modular symmetry.

The predicted neutrino mass sum,
\begin{equation}
0.059 \lesssim \sum m_\nu \lesssim 0.06\,
\mathrm{eV},
\end{equation}
satisfies the Planck bound 
$\sum m_\nu < 0.12\,$eV~\cite{Planck:2018vyg}.

The $0\nu\beta\beta$ effective mass $m_{\beta\beta}$ 
is plotted against $\sum m_\nu$ in 
Fig.~\ref{fig:mbb_m1}. The viable parameter points are highly localized,
with the total neutrino mass clustered around
$\sum m_\nu \simeq (5.7\text{--}6.0)\times10^{-2}\,\mathrm{eV}$,
while the predicted effective Majorana mass lies approximately in the
range
\begin{equation}
  1.3\times10^{-3}
  \lesssim
  m_{\beta\beta}
  \lesssim
  3.5\times10^{-3}\;\mathrm{eV}.
  \label{eq:mbb_range}
\end{equation}
The predicted values are still below the current experimental sensitivity of KamLAND-Zen \cite{KamLANDZen2016} and the projected limits of LEGEND-1000 \cite{LEGEND2021} and nEXO \cite{nEXO2018}. Nevertheless, the model predicts a narrow and well-defined region
for $m_{\beta\beta}$, which may become testable in future
high-precision $0\nu\beta\beta$ experiments.

\section{Flavoured Resonant Leptogenesis}
\label{sec:4}

We now address the question of whether the observed BAU can be explained by the current framework. It is known that the observed baryon asymmetry at low scales cannot be produced by thermal leptogenesis with hierarchical heavy neutrino masses in models with only two RHNs. In particular, studies within the scotogenic framework have shown that scenarios with two RHNs typically reside in the strong washout regime and are subject to a Davidson--Ibarra--type upper bound on the CP asymmetry. Consequently, successful thermal leptogenesis generally requires the lightest heavy neutrino mass to satisfy
$M_1 \gtrsim 10^{9\text{--}10}\,\mathrm{GeV}$, as discussed in Ref.~\cite{Hugle_2018}.
Since the present model also contains only two RHNs, this limitation applies. Therefore, we explore resonant leptogenesis, where the quasi-degenerate heavy neutrino spectrum leads to an enhancement of the CP asymmetry, to accomplish successful baryogenesis at lower mass scales. In the thermal history of the early Universe, the heavy Majorana neutrinos eventually decouple from the plasma as the temperature decreases to values comparable to their masses. The out-of-equilibrium and CP-violating decays of these states generate a non-zero lepton asymmetry. Subsequently, electroweak sphaleron transitions partially convert this lepton asymmetry into the BAU. Because the heavy-neutrino mass scale in this work lies
below $10^9\,\mathrm{GeV}$, the charged-lepton Yukawa interactions
are in thermal equilibrium throughout the epoch of asymmetry
generation. As a consequence, the three lepton flavours decohere and
must be tracked as independent dynamical degrees of freedom; the
calculation therefore requires the fully flavoured leptogenesis
treatment ~\cite{Samanta:2019yeg}.

The flavored CP asymmetry associated with the decay of the heavy neutrino $N_i$ into a lepton of flavor $\alpha$ is given by~\cite{CHAUHAN2025116908}
\begin{equation}
\epsilon_{i\alpha} =
\frac{
\Gamma(N_i \rightarrow L_\alpha \eta)
-
\Gamma(N_i \rightarrow \bar L_\alpha \eta^\dagger)
}{
\sum_\beta
\left[
\Gamma(N_i \rightarrow L_\beta \eta)
+
\Gamma(N_i \rightarrow \bar L_\beta \eta^\dagger)
\right]
}.
\end{equation}
Here $\alpha$ denotes the lepton flavour in the final state, while the
sum over $\beta$ in the denominator runs over all lepton flavours
$\beta = e,\,\mu,\,\tau$. The denominator therefore corresponds to the
total decay width of the heavy neutrino $N_i$.
Since the heavy neutrino mass matrix derived from modular invariance
naturally leads to quasi--degenerate eigenvalues, the CP asymmetry
receives a resonant enhancement. Taking into account both the mixing
and oscillation contributions, the flavored CP asymmetry can be written
as
\begin{equation}
\epsilon_{i\alpha}
=
\frac{1}{8\pi (\tilde{Y}_\nu^\dagger \tilde{Y}_\nu)_{ii}}
\sum_{j\neq i}
\mathrm{Im}
\left[
{({\tilde{Y}_{\nu})^*_{\alpha i}}}
{(\tilde{Y}_{\nu})_{\alpha j}}
(\tilde{Y}_\nu^\dagger \tilde{Y}_\nu)_{ij}
\right]
\,
\frac{M_i M_j (M_i^2-M_j^2)}
{(M_i^2-M_j^2)^2 + A^2}.
\end{equation}
where the regulator $A$ accounts for the finite decay width effects
responsible for the resonant enhancement.

The CP asymmetry receives two distinct contributions originating from
the mixing and oscillation of the nearly degenerate heavy neutrinos.
In the resonant regime both effects can become comparable and must be
taken into account simultaneously. The total CP asymmetry can therefore
be expressed as
\begin{equation}
\epsilon_{i\alpha}
=
\epsilon_{i\alpha}^{\mathrm{mix}}
+
\epsilon_{i\alpha}^{\mathrm{osc}} .
\end{equation}

The corresponding regulators entering the resonant enhancement are
given by
\begin{align}
A_{\mathrm{mix}} &= M_i \Gamma_j , \\
A_{\mathrm{osc}} &= (M_1\Gamma_1 + M_2\Gamma_2)
\sqrt{
\frac{\det(\tilde{Y}_\nu^\dagger \tilde{Y}_\nu)}
{(\tilde{Y}_\nu^\dagger \tilde{Y}_\nu)_{11}
(\tilde{Y}_\nu^\dagger \tilde{Y}_\nu)_{22}}
}.
\label{eq:Amix}
\end{align}
In the numerical analysis, we include both the mixing and oscillation 
contributions to the CP asymmetry. Where the resonant enhancement in each contribution is regulated by 
$A_{\rm mix}$ and $A_{\rm osc}$ respectively, as defined in 
Eqs.~\eqref{eq:Amix}.
The decay width of the heavy neutrino is given by
\begin{equation}
\Gamma_i =
\frac{(\tilde{Y}_\nu^\dagger \tilde{Y}_\nu)_{ii}}{8\pi} M_i
\left(1-\frac{m_\eta^2}{M_i^2}\right)^2 .
\end{equation}

The strength of washout effects in each lepton flavor is characterized
by the washout parameter
\begin{equation}
K_\alpha =
\frac{\Gamma_{D,\alpha}}{H(T=M_1)},
\end{equation}
The partial decay width of the heavy neutrino $N_1$ into the lepton flavor $\alpha$ is represented by the quantity $\Gamma_{D,\alpha}$,
\begin{equation}
\Gamma_{D,\alpha} =
\frac{| \tilde{Y}_{\alpha 1} |^2}{8\pi} M_1 ,
\end{equation}
where $\tilde{Y}_\nu = Y_\nu U_R$ denotes the Yukawa coupling matrix in the heavy-neutrino mass basis, and $H$ is the Hubble expansion rate,
\begin{equation}
H(T)=1.66\,\sqrt{g_*}\,\frac{T^2}{M_{\rm Pl}},
\end{equation}
with $g_*=106.75$ and $M_{\rm Pl}=1.22\times10^{19}\,\mathrm{GeV}$.

An approximate estimate of the baryon asymmetry generated from the
flavored lepton asymmetries is given by
\begin{equation}
Y_B =
C
\sum_\alpha
\epsilon_{1\alpha}
\,\eta(K_\alpha),
\label{eq:YB_app}
\end{equation}
where $\eta(K_\alpha)$ is the efficiency factor describing washout
effects and
\begin{equation}
C =
\frac{28}{79}
\frac{135\zeta(3)}{4\pi^4 g_*}
\end{equation}
accounts for sphaleron conversion and entropy dilution.

In order to describe the asymmetry evolution with improved accuracy, we analyze the flavor-dependent Boltzmann equations. By defining the dimensionless quantity $z = M_1/T$, one can express the evolution of both the heavy neutrino population and the individual lepton asymmetries as follows~\cite{Samanta:2019yeg}.
\begin{align}
\frac{dY_{N_1}}{dz} &=
- D(z)\left(Y_{N_1}-Y_{N_1}^{ eq}\right), \\
\frac{dY_{\Delta_\alpha}}{dz} &=
\epsilon_{1\alpha}
D(z)\left(Y_{N_1}-Y_{N_1}^{ eq}\right)
-
W_\alpha(z)Y_{\Delta_\alpha}.
\label{eq:boltzz}
\end{align}

The equilibrium abundance of the heavy neutrino is
\begin{equation}
Y_{N_1}^{ eq}(z)
=
\frac{45}{4\pi^4 g_*}
z^2 K_2(z),
\end{equation}
where $K_n$ denotes the modified Bessel functions.

The quantities governing the decay and washout processes in the Boltzmann equations are defined as
\begin{align}
D(z) &= K_T\,z\frac{K_1(z)}{K_2(z)}, \\
W_\alpha(z) &= \frac{1}{2}K_\alpha z^3 K_1(z),
\end{align}
with $K_T=\sum_\alpha K_\alpha$.

The baryon asymmetry generated during the evolution is obtained from
the total lepton asymmetry through sphaleron conversion,
\begin{equation}
Y_B(z) = \frac{28}{79}
\sum_\alpha Y_{\Delta_\alpha}(z).
\end{equation}
\subsection{Numerical Analysis}

We turn now to the leptogenesis analysis within the modular 
$S_4$ scotogenic framework. The parameter points identified 
in Sec.~\ref{sec:3} as consistent with all neutrino oscillation 
data serve as the input for this computation. For each such 
point, we evaluate the flavour-dependent CP asymmetries 
arising from heavy Majorana neutrino decays and determine 
the resulting baryon yield generated through the 
out-of-equilibrium decay mechanism.
 
\begin{figure}[t]
\centering
\begin{minipage}{0.95\textwidth}
\centering
\includegraphics[width=0.49\textwidth]{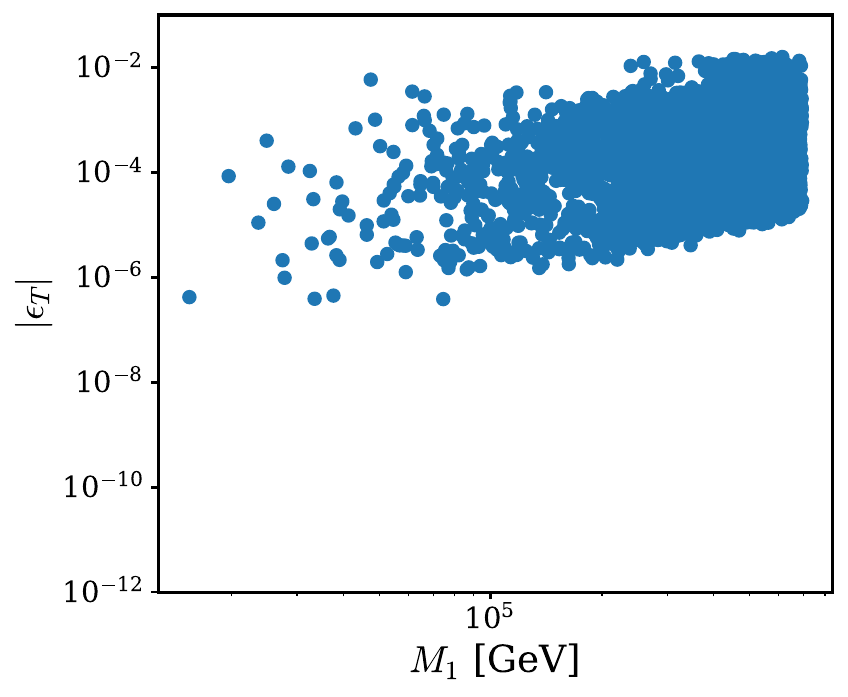}
\includegraphics[width=0.49\textwidth]{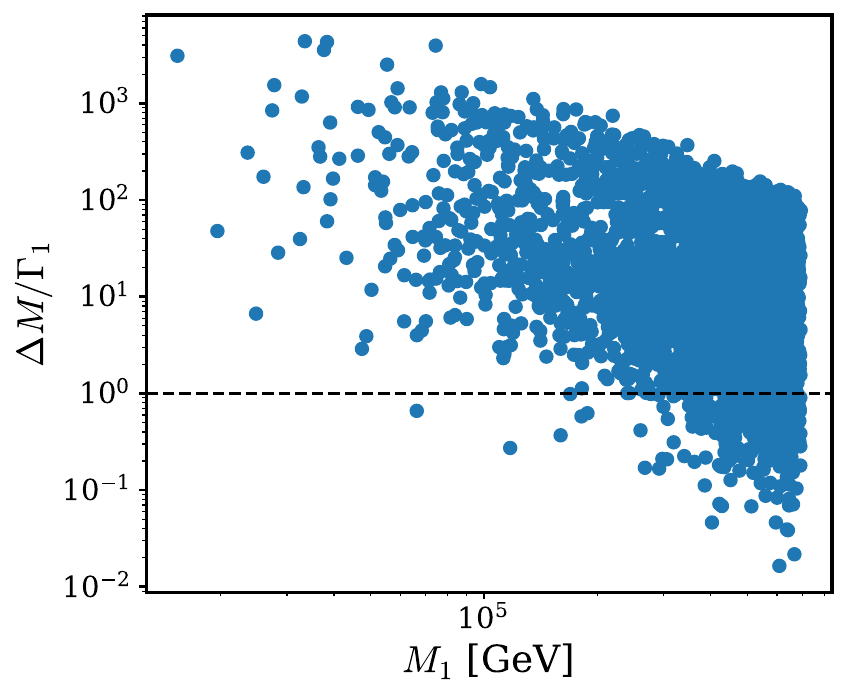}
\end{minipage}

\vspace{0.3cm}

\begin{minipage}{0.95\textwidth}
\centering
\includegraphics[width=0.49\textwidth]{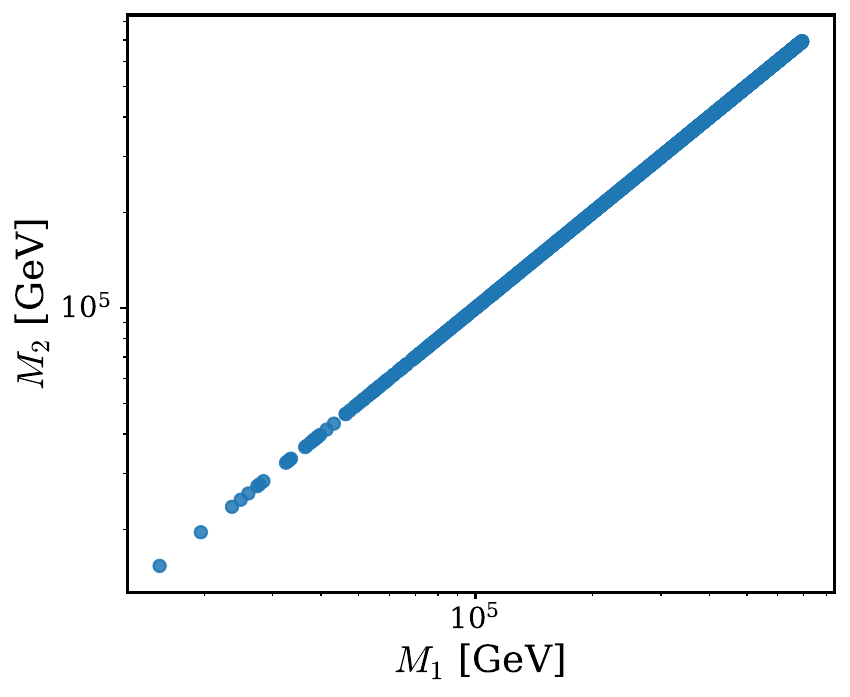}
\end{minipage}

\caption{Top-left: Total CP asymmetry $|\epsilon_T|$ 
vs.\ $M_1$. Top-right: Resonance parameter 
$\Delta M/\Gamma_1$ vs.\ $M_1$; the dashed line 
marks $\Delta M/\Gamma_1 = 1$. Bottom: $M_2$ 
vs.\ $M_1$, confirming the quasi-degenerate 
spectrum across the viable parameter space.}
\label{fig:eps1}
\end{figure}

The behaviour of the CP asymmetry and the resonance condition
as functions of the heavy-neutrino mass scale is illustrated in
Fig.~\ref{fig:eps1}.
The left panel displays the magnitude of the total CP asymmetry
$|\epsilon_T| = |\epsilon_e+\epsilon_\mu+\epsilon_\tau|$
as a function of the lightest heavy neutrino mass $M_1$.
The phenomenologically acceptable points concentrated in the range $10^{-7} \lesssim |\epsilon_T| \lesssim 10^{-2}$.
This wide spread is a direct consequence of the resonant
amplification that arises because the modular structure of $M_R$
forces the two heavy neutrino masses to be nearly degenerate. In the present analysis, we employ an effective single-species treatment 
and track only the evolution of $N_1$. Since $M_1 \approx M_2$, both 
heavy neutrinos share essentially identical equilibrium distributions and 
decay rates, so their number densities evolve in an almost identical 
fashion. The CP asymmetry is nevertheless computed using the full resonant 
expressions involving both quasi-degenerate heavy neutrinos $N_1$ and 
$N_2$, including the mixing and oscillation contributions as defined in 
Eqs.~\eqref{eq:Amix}. This approximation therefore 
captures the dominant resonant enhancement while keeping the Boltzmann 
analysis tractable.
 
The right panel of Fig.~\ref{fig:eps1} shows how the ratio 
$\Delta M/\Gamma_1$ varies with the heavy neutrino mass $M_1$. 
Here $\Delta M = |M_2 - M_1|$ is the mass gap between the two 
nearly degenerate heavy neutrinos, and $\Gamma_1$ denotes the 
total decay width of the lighter eigenstate. For resonant 
leptogenesis to operate efficiently, this ratio must satisfy 
$\Delta M/\Gamma_1 \sim \mathcal{O}(1)$, meaning the splitting 
and the decay width are of comparable magnitude. When this 
condition holds, the self-energy diagram contributes a 
resonantly amplified correction to the CP asymmetry, 
substantially enhancing the net lepton number generated 
per decay.

The lower panel of Fig.~\ref{fig:eps1} displays the correlation between $M_1$ and $M_2$. The two eigenvalues track each other extremely closely, confirming
that the quasi-degenerate spectrum persists throughout the
acceptable parameter space.
This degeneracy is not an accident of parameter tuning but a
structural feature inherited from the modular form of $M_R$,
as discussed in Sec.~\ref{sec:2}.

\begin{figure}[t]
\centering
\includegraphics[width=0.75\textwidth]{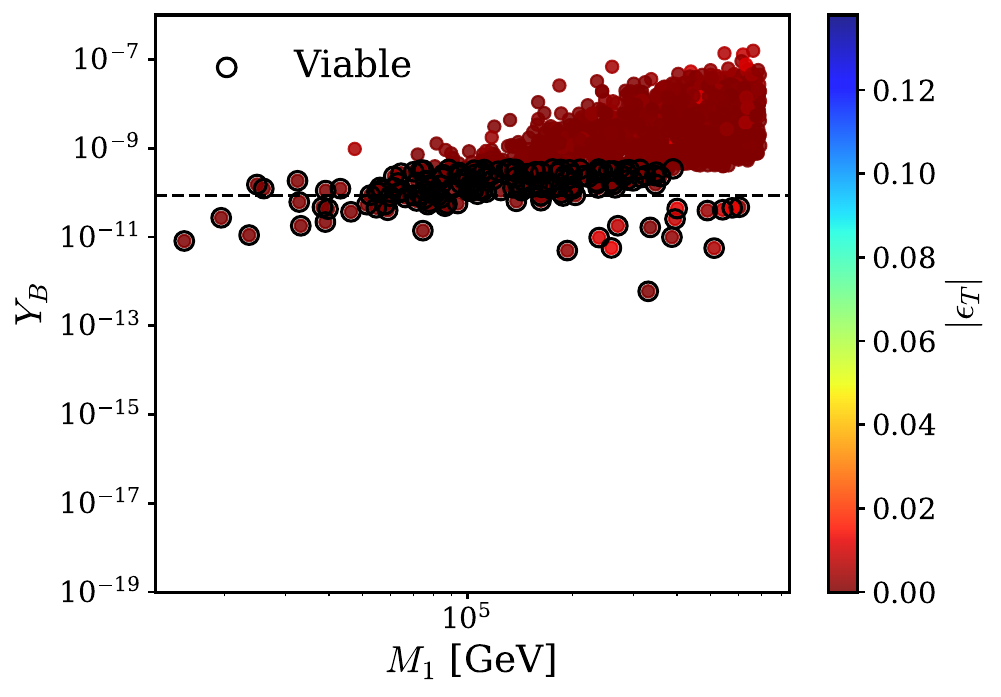}
\caption{
Correlation between the baryon asymmetry and the heavy-neutrino mass $M_1$. The dashed line marks the observed value of the baryon asymmetry. Circled points correspond to phenomenologically viable solutions, and the colour coding indicates the absolute value of the total CP asymmetry $|\epsilon_T|$.
}
\label{fig:YBM1}
\end{figure}

The final baryon asymmetry is shown in Fig.~\ref{fig:YBM1} as a
function of $M_1$.
Points that reproduce the cosmological measurement
$Y_B^{ obs}\simeq 8.6\times10^{-11}$~\cite{Planck:2018vyg}
(dashed horizontal line) are highlighted, and the colour axis
encodes $|\epsilon_T|$.
The observed asymmetry is recovered for $M_1\sim10^5\,\mathrm{GeV}$,
and a clear positive trend is visible: points with a higher
CP asymmetry yield a proportionally larger baryon asymmetry,
in line with the analytic estimate of Eq.(~\ref{eq:YB_app}). We note that the viable points identified in Fig.~\ref{fig:YBM1}  are obtained
using the semi-analytic estimate of Eq.(~\ref{eq:YB_app}), where $Y_B$ scales linearly with $\epsilon_{1\alpha}$. The full numerical
integration of the Boltzmann equations, which consistently accounts
for the continuous interplay between CP-asymmetry generation and
strong washout effects, generally modifies the final value of
$Y_B$ relative to the analytic approximation, as illustrated by the
benchmark solutions discussed below.

\begin{table}[t]
\centering
\caption{
Representative benchmark points used for solving the flavored
Boltzmann equations.
}
\begin{tabular}{c c c}
\hline\hline
Parameter & Benchmark 1 & Benchmark 2 \\
\hline
$M_1$ [GeV] & $7.53946\times10^5$ & $7.33497\times10^5$ \\
$M_2$ [GeV] & $7.53947\times10^5$ & $7.33498\times10^5$ \\
\hline
$\epsilon_e$ & $-1.85482\times10^{-4}$ & $-2.26363\times10^{-3}$ \\
$\epsilon_\mu$ & $6.15317\times10^{-5}$ & $5.99216\times10^{-4}$ \\
$\epsilon_\tau$ & $7.48281\times10^{-5}$ & $7.63612\times10^{-4}$ \\
\hline
$K_e$ & $8.67334\times10^{2}$ & $2.07489\times10^{4}$ \\
$K_\mu$ & $6.16650\times10^{1}$ & $1.19754\times10^{3}$ \\
$K_\tau$ & $3.35990\times10^{2}$ & $6.65183\times10^{3}$ \\
\hline
$Y_B$ & $3.20448\times10^{-10}$ & $1.15114\times10^{-10}$ \\
\hline\hline
\end{tabular}

\label{tab:benchmarks}
\end{table}
\begin{figure}[htbp]
\centering
\includegraphics[width=1.0\textwidth]{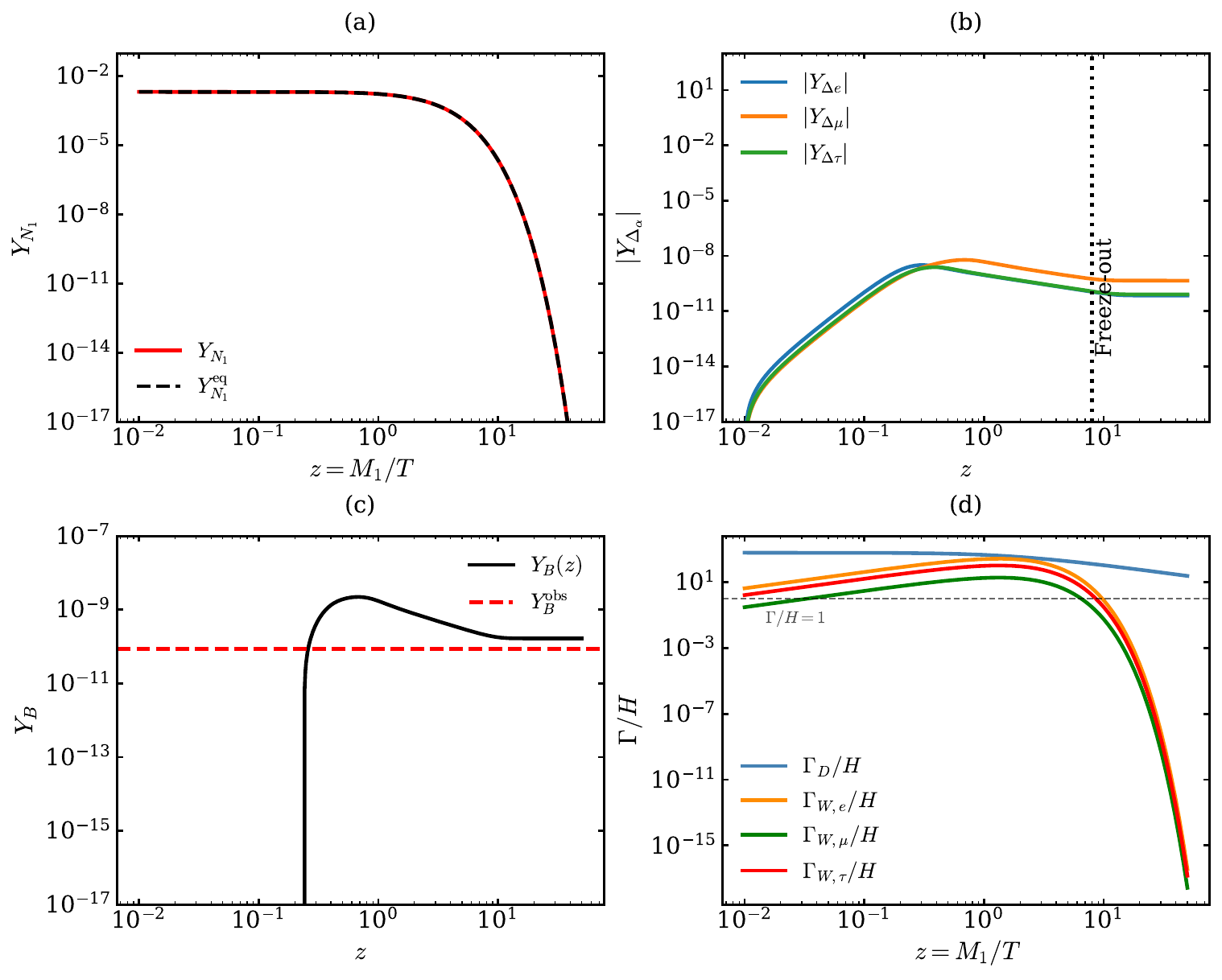}
\caption{Boltzmann evolution for Benchmark 1.}
\label{fig:BP1}
\end{figure}

\begin{figure}[h]
\centering
\includegraphics[width=1.0\textwidth]{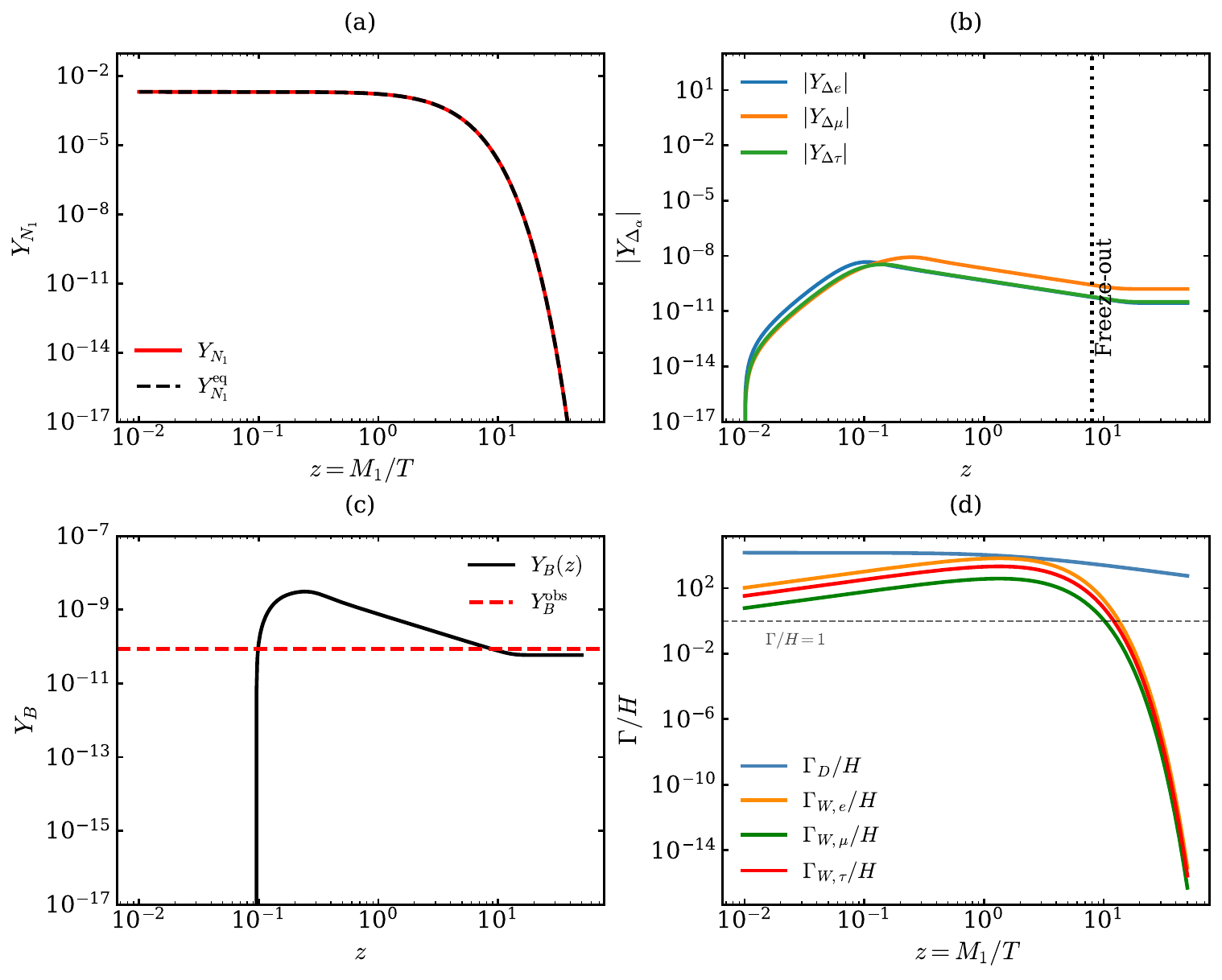}
\caption{Boltzmann evolution for Benchmark 2.}
\label{fig:BP2}
\end{figure}

To examine the dynamics of asymmetry generation in detail, we
integrate the full set of three-flavour Boltzmann equations in
Eqs.~(\ref{eq:boltzz}). The model parameters for two representative benchmark configurations are listed in Table~\ref{tab:benchmarks}. The Boltzmann evolution for both benchmarks is presented in
Figs.~\ref{fig:BP1} and~\ref{fig:BP2}, each comprising four panels
that together capture the complete asymmetry-generation history.
The evolution of the heavy neutrino abundance $Y_{N_1}$ and the equilibrium distribution $Y_{N_1}^{eq}$ as a function of $z=M_1/T$ are displayed in the upper left panel.
At early times ($z\ll1$), the heavy neutrinos remain in thermal
equilibrium with the plasma and therefore closely follow the
equilibrium abundance. As the temperature decreases to values comparable to the heavy neutrino masses, the interaction rate becomes smaller than the Hubble expansion rate. Consequently, the heavy neutrinos fall out of thermal equilibrium and their abundance departs from the equilibrium distribution.
The heavy neutrinos subsequently decay, providing the necessary
out--of--equilibrium condition for the generation of a net lepton
asymmetry.

The upper right panel displays the evolution of the flavored lepton
asymmetries $|Y_{\Delta e}|$, $|Y_{\Delta \mu}|$, and
$|Y_{\Delta \tau}|$. The asymmetries are generated through
CP--violating decays of the heavy neutrinos and are partially erased
by washout processes. Because the washout parameters differ among the
three flavors, the individual asymmetries evolve differently during
the thermal history. In the first benchmark point the hierarchy
$|Y_{\Delta\mu}| > |Y_{\Delta e}|\approx |Y_{\Delta\tau}|$ is observed,
indicating that the muon flavor provides the dominant contribution
to the total asymmetry.
Similarly in the second benchmark point exhibits the hierarchy
$|Y_{\Delta\mu}| \gg |Y_{\Delta e}| \approx |Y_{\Delta\tau}|$, where
the muon asymmetry again dominates while the electron and tau
asymmetries remain comparable but subdominant. These differences
reflect the flavor dependence of the washout parameters and highlight
the importance of treating leptogenesis in a fully flavored framework. The total CP asymmetry
$\epsilon_T\simeq-4.91\times10^{-5}$ is negative for
Benchmark~1, while the final baryon asymmetry stabilises at the
positive value $Y_B\simeq3.20\times10^{-10}$. In the fully flavoured regime, however, the baryon asymmetry is
not determined by the unweighted total CP asymmetry
$\epsilon_T=\sum_\alpha \epsilon_{1\alpha}$, but by the weighted
combination $\sum_\alpha \epsilon_{1\alpha}\eta(K_\alpha)$
appearing in Eq.(~\ref{eq:YB_app}). This apparent sign
reversal originates entirely from the flavour-dependent washout
effects. In particular, the efficiency factor
$\eta(K_\alpha)$ strongly suppresses the electron flavour
contribution since $K_e\simeq867$ corresponds to the largest
washout parameter, whereas the muon flavour contribution survives
most efficiently due to the comparatively smaller washout
parameter $K_\mu\simeq62$. Consequently, the weighted flavour sum
$\sum_\alpha \epsilon_{1\alpha}\eta(K_\alpha)$ becomes dominated
by the positive muon and tau contributions despite the negative
total asymmetry $\epsilon_T$, ultimately yielding a positive
final baryon asymmetry through Eq.(~\ref{eq:YB_app}).
The vertical dashed line indicates the freeze--out epoch at which the
washout processes become inefficient and the asymmetries stop evolving
significantly.

The lower left panel shows the evolution of the baryon asymmetry
$Y_B(z)$ obtained after sphaleron conversion of the total lepton
asymmetry. The baryon asymmetry begins to grow once the heavy
neutrinos start decaying out of equilibrium at
$z\sim\mathcal{O}(1)$. In the first benchmark point, the asymmetry
grows rapidly around $z\sim1$, overshoots the observed value, and
is then partially washed back down before stabilizing at
$Y_B\simeq3.20\times10^{-10}$, which lies a factor of
$\sim3.7$ above
$Y_B^{ obs}\simeq8.6\times10^{-11}$. The observed overshoot is a consequence of the sizable CP asymmetry $\epsilon_T$ together with the temporary dominance of asymmetry generation over washout processes near $z \sim 1$. During this stage, the baryon asymmetry rises above the observed value, $Y_B^{\mathrm{obs}}$, before being partially reduced by washout effects and eventually settling to its final value. The incomplete washout,
particularly due to the comparatively weaker muon washout channel
with $K_\mu\simeq62$, prevents the generated asymmetry from being
fully erased and leaves a residual $Y_B$ above the observed value.
In the second benchmark point, the asymmetry evolves more
gradually, rising smoothly and stabilizing at
$Y_B\simeq1.15\times10^{-10}$, a factor of $\sim1.3$ above
$Y_B^{ obs}$. The more gradual evolution in Benchmark~2 is a
consequence of the larger washout parameters $K_\alpha$ relative
to Benchmark~1, which more efficiently suppress the asymmetry and
keep the inverse decay and washout processes active over a longer
interval in $z$, leading to a smoother departure from equilibrium
and a more gradual asymmetry build-up. The dashed horizontal line
in panel~(c) marks $Y_B^{ obs}$. Both benchmark trajectories
stabilize above the observed value, with Benchmark~2 approaching
it more closely. The observed baryon asymmetry is reproduced by
nearby points in the parameter space as shown in Fig.~7, where the
analytic estimate identifies viable points landing on
$Y_B^{ obs}$. The deviation of the benchmark points from the
exact observed value reflects the dynamical strong washout
continuously erasing the generated asymmetry during the Boltzmann
evolution in a way that the analytic efficiency factor
$\eta(K_\alpha)$ of Eq.~(4.10) does not fully capture.

Finally, the lower right panel displays the decay rate
$\Gamma_D/H$ and the three flavoured washout rates
$\Gamma_{W,\alpha}/H$ as functions of $z$. At early times
($z\ll1$), all rates begin well above the Hubble rate with
$\Gamma/H\gg1$, confirming that the system initialises in full
thermal equilibrium. At this stage any generated asymmetry is
efficiently erased since the interaction rates are too fast to
sustain a departure from equilibrium. A necessary Sakharov condition for baryogenesis is the departure from thermal equilibrium. In the present scenario, this condition is realized when the relevant interaction rates become slower than the Hubble expansion rate at approximately $z \sim \mathcal{O}(1)$. For Benchmark~1, the decay rate
$\Gamma_D/H$ crosses unity at $z\sim\mathcal{O}(1)$, triggering
the onset of asymmetry generation. The three flavoured washout
rates cross the threshold at different values of $z$ due to the
hierarchy among the washout parameters
$K_e\simeq867\gg K_\tau\simeq336\gg K_\mu\simeq62$: the muon
washout freezes out earliest since its rate crosses
$\Gamma/H=1$ at the smallest $z$, followed by the tau and
electron channels respectively, allowing the muon flavour
asymmetry to survive most effectively. For Benchmark~2, the same
qualitative ordering holds but all washout rates remain above
unity for longer owing to the significantly larger washout
parameters
$K_e\simeq2.07\times10^4$,
$K_\mu\simeq1.20\times10^3$,
$K_\tau\simeq6.65\times10^3$,
resulting in stronger overall suppression of all flavoured
asymmetries. It is precisely this flavour-by-flavour freeze-out
structure, with each lepton flavour decoupling at a different
epoch, that makes the fully flavoured three-flavour Boltzmann
treatment indispensable --- a single-flavour approximation would
fail to capture these differences and yield an unreliable
prediction for $Y_B$.

This result demonstrates that the unflavoured approximation
fails qualitatively for the present model in two important ways.
First, using the total CP asymmetry
$\epsilon_T\simeq-4.91\times10^{-5}$ within an unflavoured
framework would predict $Y_B<0$, thereby yielding the wrong sign
for the baryon asymmetry. Second, even if one artificially
replaces $\epsilon_T$ by $|\epsilon_T|$ to recover a positive
asymmetry, the predicted magnitude would still be incorrect
because a single unflavoured efficiency factor $\eta(K_T)$ cannot
capture the flavour-dependent washout dynamics. In particular,
the positive muon contribution survives washout much more
efficiently due to the comparatively smaller washout parameter
$K_\mu\simeq62$, whereas the negative electron contribution is
strongly suppressed by the much larger washout parameter
$K_e\simeq867$. Only the fully flavoured treatment, which tracks
each contribution $\epsilon_{1\alpha}\eta(K_\alpha)$
independently, correctly reproduces both the sign and magnitude
of the final baryon asymmetry $Y_B$.

\section{Conclusion}
\label{sec:6}

We have built and examined a modular-invariant scotogenic model in which the finite modular group $S_4$ controls the flavor symmetry. The framework's fundamental organizational principle is that all Yukawa couplings come from modular forms, which are holomorphic functions of a single complex modulus $\tau$. This means that a single parameter, rather than a collection of flavon VEVs, determines the entire lepton sector's flavor structure. The 
scotogenic mechanism for radiative neutrino mass generation is 
embedded within this modular structure, yielding a model that 
simultaneously addresses neutrino masses, leptonic mixing, and the 
BAU within a unified 
and economical theoretical framework.

A comprehensive scan over the modulus $\tau$ and the Yukawa parameter 
space was carried out, retaining only solutions consistent with all 
five neutrino oscillation observables at the $3\sigma$ level from NuFIT 
5.2. The model proves to be highly predictive. The three leptonic mixing angles are confined to the intervals
$\sin^2\theta_{12}\simeq 0.290$--$0.329$,
$\sin^2\theta_{23}\simeq 0.480$--$0.584$, and
$\sin^2\theta_{13}\simeq 0.0210$--$0.0237$,
all in agreement with current global-fit constraints.
The mass-squared differences lie in the ranges
$\Delta m^2_{21}\simeq (7.32$--$7.57)\times10^{-5}\,\mathrm{eV}^2$
and
$\Delta m^2_{31}\simeq (2.50$--$2.53)\times10^{-3}\,\mathrm{eV}^2$,
consistent with experimental determinations.

The presence of exactly two RHNs forces the light 
neutrino mass matrix to have rank two, which has two immediate 
physical consequences: a single massless neutrino eigenstate at 
leading order, and the selection of the normal mass ordering as the 
only viable option. The resulting neutrino mass sum is tightly bounded 
to $0.059 \lesssim \Sigma m_\nu \lesssim 0.06\ \mathrm{eV}$, satisfying the cosmological upper limit 
and falling within the sensitivity of forthcoming large-scale 
structure surveys. The effective Majorana mass relevant for $0\nu\beta\beta$ is expected to fall inside the range
$1.3 \times 10^{-3} \lesssim m_{\beta\beta} \lesssim 3.5 \times 10^{-3}\,\mathrm{eV}$.
These values are below the projected sensitivities of upcoming experiments such as LEGEND-1000 and nEXO, making their detection challenging in the near future.

A structurally important feature of the model is that the modular 
form of the right-handed Majorana mass matrix $M_R$ naturally produces 
a quasi-degenerate heavy neutrino spectrum without any parameter 
tuning. This near-degeneracy serves as the dynamical engine for 
resonant leptogenesis by resonantly amplifying the CP asymmetry in 
heavy neutrino decays. Our analysis shows that the model generates baryon asymmetries of the correct order of magnitude for $M_1 \sim 10^5\ \mathrm{GeV}$, with the observed value $Y_B^{\mathrm{obs}} \simeq 8.6 \times 10^{-11}$ reproduced within viable regions of the parameter space identified through the analytic estimate. The corresponding CP asymmetries span the range $|\epsilon_T| \sim 10^{-7}\text{--}10^{-2}$. The complete asymmetry generation dynamics were investigated by integrating the full three-flavour Boltzmann equation system for two representative benchmark configurations. The resulting baryon yields are of the correct order of magnitude and illustrate the crucial role of flavour-dependent washout effects in resonant leptogenesis. 
A key finding is that the three lepton flavours decouple at markedly 
different epochs due to the hierarchy among the washout parameters 
$K_\alpha$, making the single-flavour approximation qualitatively unreliable 
and the fully flavoured treatment indispensable.

In summary, the modular $S_4$ scotogenic framework provides a predictive and economical description of neutrino masses, leptonic mixing, and baryogenesis within a unified theoretical setting controlled by the modulus $\tau$. The model yields characteristic correlations among neutrino observables that can be tested in future oscillation experiments, cosmological measurements of the neutrino mass sum, and $0\nu\beta\beta$ searches. Further exploration of the DM sector, including its detection prospects and collider signatures, remains an interesting direction for future investigation.

\bibliographystyle{unsrt}

\end{document}